\documentclass[preprint,showpacs,preprintnumbers,amsmath,amssymb]{revtex4}

\usepackage{graphicx}% Include figure files
\usepackage{dcolumn}% Align table columns on decimal point
\usepackage{bm}% bold math

\begin{document}

\preprint{APS/123-QED}

\title{Bethe-Salpeter approach for relativistic positronium in
a strong magnetic field}% Force line breaks with \\

\author{A.E. Shabad}

 \email{shabad@lpi.ru}
\affiliation{%
P.N. Lebedev Physics Institure, Moscow, Russia\\}

\author{V.V. Usov}
 \email{fnusov@wicc.weizmann.ac.il}
\affiliation{Center for Astrophysics, Weizmann Institute of
Science, Rehovot 76100, Israel\\}

\date{\today}% It is always \today, today,
             %  but any date may be explicitly specified

\begin{abstract}
We study the electron-positron system in a strong magnetic field
using the differential Bethe-Salpeter equation in the ladder
approximation. We derive the fully relativistic two-dimensional
form that the four-dimensional Bethe-Salpeter equation takes in
the limit of asymptotically strong constant and homogeneous
magnetic field. An ultimate value for the magnetic field is
determined, which provides the full compensation of the
positronium rest mass by the binding energy in the maximum
symmetry state and vanishing of the energy gap separating the
electron-positron system from the vacuum. The compensation becomes
possible owing to the falling to the center phenomenon that occurs
in a strong magnetic field because of the dimensional reduction.
The  solution of the Bethe-Salpeter equation corresponding to the
vanishing energy-momentum of the electron-positron system is
obtained.

\end{abstract}

% PACS, the Physics and
%Astronomy
                             % Classification Scheme.
%\keywords{Suggested keywords}%Use showkeys class option if keyword
                              %display desired
\maketitle

\section{Introduction}

It is well known that the structure of atoms (positronium
included) is drastically modified by a magnetic field ${\bf B}$ if
the field strength $B=|{\bf B}|$ exceeds the characteristic atomic
value $B_{\rm a}=m^2e^3c/\hbar^3\simeq 2.35\times 10^9$~G
\cite{loudon,Lai01}. In a strong magnetic field ($B\gg B_{\rm a}$)
the usual perturbative treatment of the magnetic effects (such as
Zeeman splitting of atomic energy levels) is not applicable, and
instead, the Coulomb forces act as a perturbation to the magnetic
forces. For positronium in such a field the characteristic size of
the electron and positron across ${\bf B}$ is the Larmour radius
\begin{eqnarray}\label{LB}
L_B=(eB)^{-1/2}=a_0 (B_{\rm a}/B)^{1/2}
\end{eqnarray}
and decreases with increase of the field strength, where $a_0$ is
the Bohr radius, $a_0=(\alpha m)^{-1}$, $\alpha=1/137$.
Henceforth, we set $\hbar=c=1$ and refer to the Heaviside-Lorentz
system of units, where the fine structure constant
$\alpha=e^2/4\pi$.

The properties of positronium in a strong magnetic field ($B\gg
B_{\rm a}$) are interesting for astrophysics because such fields
are observed now for several kinds of astronomical compact objects
(pulsars, powerful X-ray sources, soft gamma-ray repeaters, etc.).
Besides, some of these objects are the sources of
electron-positron pairs produced in their vicinities by various
mechanisms \cite{St71}. At least a part of these pairs may be
bound. For instance, at the surface of radio pulsars identified
with rotation-powered neutron stars the field strength is in the
range from $\sim 10^9$ G to $\sim 10^{14}$ G \cite{M03}. A common
point of all available models of pulsars is that electron-positron
pairs dominate in the magnetosphere plasma \cite{M91}. These are
formed by the single-photon production process in a strong
magnetic field, $\gamma + B\rightarrow e^+ +e^- +B$. If the field
strength is higher that $\sim 4\times 10^{12}$ G the  pairs
created are mainly bound \cite{SU85}.
Much more
intense magnetic 
fields have been conjectured to be involved in several astrophysical 
phenomena. For instance,
superconductive cosmic strings, if 
they exist, may have magnetic
fields up to $\sim 10^{47}-10^{48}$ G 
in their vicinities
\cite{W85}. Electron-positron pairs may be 
produced near such strings \cite{BHV01}. 

In  magnetic fields larger than
$B_{\rm a}$, the Coulomb force
becomes more effective in binding the positronium
because  the charged constituents are confined to the lowest
Landau level and hence to a narrow region stretching along the
magnetic field ($L_B\ll a_0$). Notwithstanding this effect, the
binding energy of positronium $\Delta E$ is still very small in
comparison with the rest mass, $\Delta E\ll
2m$, even for the fields larger than  Schwinger's critical value
$B_0=m^2/e\simeq 4.4\times 10^{13}$ G, , i.e., the positronium
remains an internally nonrelativistic system.
%($B\lesssim B_{\rm a}$),even
%notwithstanding that % In a strong magnetic field ($B\gg B_{\rm a}$)
The binding energy of the ground state, as calculated
nonrelativistically,
\begin{eqnarray}\label{deltaEB}
\Delta E \simeq \frac{m\alpha^2}{4}\left(\ln
\frac{B}{B_0}\right)^2\,,
\end{eqnarray}increases with increase of $B$,
and the relativistic effects, for extremely huge fields, should be
expected to become essential. The unrestricted growth of the
binding energy (\ref{deltaEB}) with the magnetic field is a
manifestation of the fact that the Coulomb attraction force
becomes supercritical in the one-dimensional Schr$\ddot{\rm
o}$dinger equation, to which the nonrelativistic problem is
reduced in the high-field limit \cite{loudon}, and the
falling-to-the-center phenomenon  occurs in the limit $B=\infty$.

Relativistic properties of positronium in a strong magnetic
field were studied basing on the  %by different methods, including the
Bethe-Salpeter equation \cite{leinson1,ShUs}. %The (relativistic) motion of the
%positronium as a whole was considered in ,
The nontrivial energy dependence upon the transversal
(pseudo)momentum component of the center-of-mass was found in
\cite{ShUs,leinson2}. %(for the hydrogen atom see \cite{lai}).
Although the Bethe-Salpeter equation is fully relativistic, it was
 used within the customary "equal-time" approximation that disregards the
retardation effects, so that the relative motion of the electron
and positron is treated in a nonrelativistic way. In this way the
behavior (\ref{deltaEB}) is reproduced for the ground state
\cite{leinson1} - \cite{leinson2}. A completely relativistic
solution for positronium in a strong magnetic field remains
unknown. In this paper we study the positronium in an
asymptotically strong magnetic field with not only the
center-of-mass  motion considered relativistically, but also the
relative motion of its constituents. We point the ultimate
value of the magnetic field guaranteeing such deepening of the
positronium energy level that is sufficient to compensate for the
whole rest mass $2m$ of it.

To this end, in Section~II we derive  the fully relativistic - in
two-dimensional Minkowsky  space - form  that the differential
Bethe-Salpeter equation in the ladder approximation takes for the
positronium  when the magnetic field tends to infinity. This
equation is efficient already for $B\gg B_{\rm a}$. 
We also include a moderate external electric field parallel to
$\bf B$ into this equation. In Section~III the ultra-relativistic
solution of maximum symmetry  is found to the equation derived in
Section~II corresponding to the vanishing total energy-momentum of
the positronium. The falling-to-the-center phenomenon \cite{QM}
characteristic of the two-dimensional equation of Section~II for
every positive value of the fine structure constant
\cite{newfootnote} is exploited for establishing the possibility
that the zero energy point may belong to the spectrum, provided
the magnetic field is sufficiently large. The origin of the
falling to the center is in the ultraviolet singularity of the
photon propagator. The effects of the mass radiative corrections
and of the vacuum polarization are also considered. In
concluding Section~IV the results are summarized.

\section{Two-dimensional Bethe-Salpeter equation for positronium
in an asymptotically strong magnetic field}

The view \cite{loudon} that charged particles in a strong constant
magnetic field are confined to the lowest Landau level and behave
effectively as if they possess only one spacial degree of freedom
- the one along the magnetic field - is widely accepted. Moreover,
a conjecture exists \cite{skobelev} that the Feynman rules in the
high magnetic field limit may be directly served by
two-dimensional (one space + one time)  form of electron
propagators. As applied to the Bethe-Salpeter equation, the
dimensional reduction in high magnetic field was considered in
\cite{leinson1,ShUs}. In these references the well-known
simultaneous approximation to the Bethe-Salpeter equation taken in
the integral form was exploited, appropriate for nonrelativistic
treatment of the relative motion of the two charged particles.
Once we shall  in the next Section be interested in the
ultrarelativistic regime, we reject from using this approximation,
and find it convenient to deal only with the differential form of
the Bethe-Salpeter equation.

The electron-positron bound state is described by the
Bethe-Salpeter amplitude (wave function)
{\Large$\chi$}$_{\lambda,\beta}(x^{\rm e},x^{\rm p})$ subject to
the fully relativistic equation (e.g., \cite{schweber}), which in the
ladder approximation in a magnetic field may be written as
\begin{eqnarray}\label{equation}[{\rm i}
\hat{\partial}^{\rm e}-m+e\hat{A}(x^{\rm e})] _{\lambda \beta}
[{\rm i}\hat{\partial}^{\rm p}-m-e\hat{A}(x^{\rm p})]_{\mu\nu}
\begin{tabular}{c}{\Large$\chi$}$_
{\beta\nu}(x^{\rm e},x^{\rm p})$\end{tabular}\nonumber\\
=-{\rm i} 8\pi\alpha D^{i j}(x^{\rm e}-x^{\rm
p})[\gamma_i]_{\lambda\beta}[\gamma_j]_{\mu\nu}
\begin{tabular}{c}{\Large$\chi$}$_
{\beta\nu}(x^{\rm e},x^{\rm p}).$\end{tabular}\; \end{eqnarray}
Here $x^{\rm e},x^{\rm p}$ are the electron and positron
4-coordinates, $D^{ij}(x^{{\rm e}}-x^{{\rm p}})$ is the photon
propagator, and we have explicitly written the spinor indices
$\lambda,~\beta,~\mu,~\nu=1,2,3,4.$ The metrics in the Minkowsky
space is $\;{\rm diag}\; g_{ij}=(1,-1,-1,-1), ~i,j=0,1,2,3$. The
derivatives \begin{eqnarray}\label{derivatives}
\hat{\partial}=\partial^j\gamma_j=\partial^0\gamma_0+
\partial^k\gamma_k=\gamma_0\frac{\partial}{\partial x_0}+
\gamma_k\frac\partial{\partial
x_k}, \quad k=1,2,3, \end{eqnarray} act on $x^{\rm e}$ or $x^{\rm
p}$ as indicated by the superscripts, and
$\hat{A}=A_0\gamma_0-A_k\gamma_k$.

We consider the ladder approximation with the photon propagator
taken in the Feynman gauge. With other gauges this approximation
corresponds to summation of diagrams other than the ladder ones in
agreement with the well-known fact that the ladder approximation
is not gauge-invariant.

We refer to, if needed, the so called spinor representation of the
Dirac $\gamma$-matrices in the block form
\begin{eqnarray}\label{gamma}
\gamma_0=\left(\begin{tabular}{cc}0&I\\I&0\end{tabular}\right),\qquad
\gamma_k=\left(\begin{tabular}{cc}0&$-\sigma_k$\\$\sigma_k$&0\end{tabular}
\right),\end{eqnarray} $\sigma_k$ are the Pauli matrices:
 \begin{eqnarray}\label{pauli} \sigma_1=\left(\begin{tabular}{cc} 0
&1\\1 &
 0\\
\end{tabular}\right),\quad {\rm i}\sigma_2=\left(\begin{tabular}{cc} 0
&1\\-1&
0\\
\end{tabular}\right),\quad \sigma_3=\left(\begin{tabular}{cc} 1& 0\\0&
-1\\\end{tabular}\right),\end{eqnarray} $m$ is the electron mass,
$e$ the absolute value of its charge, $e=2\sqrt{\pi\alpha}$.  The
vector potential of the constant and homogeneous magnetic field
$B$, directed along the axis 3 ($ B_3=B,~ B_{1,2}=0$), is chosen
in the asymmetric gauge
\begin{eqnarray}\label{asym}A_1(x)=-Bx_2,\;\;\;A_{0,2,3}(x)=0.
\end{eqnarray}
With this choice, the translational invariance along the
directions 0,1,3 holds.

Solutions to Eq. (\ref{equation}) may be represented in the form
 \begin{eqnarray}\label{trans}
\begin{tabular}{c}
 {\Large$
 \chi$}$(x^{{\rm e}},x^{{\rm p}})=$\end{tabular}
 %\nonumber\\
\begin{tabular}{c}
 {\Large$
 \eta$}$(x_0^{{\rm e}}-x_0^{{\rm p}},x_3^{{\rm e}}-x_3^{{\rm
p}},x^{{\rm e}}_
 {1,2},x^{{\rm p}}_{1,2})$
\end{tabular}
 \exp\{\frac{\rm i}{2}[P_0(x_0^{{\rm e}}+
 x^{{\rm p}}_0)-P_3(x_3^{{\rm e}}+x^{{\rm p}}
 _3)]\}, \end{eqnarray}
 where $P_{0,3}$ are the center-of-mass 4-momentum components of
 the longitudinal motion,
 that express the translational invariance along the
 longitudinal directions (0,3).
 Denoting the differences $x_0^{\rm e}-x_0^{\rm p}=t,$
 $x_3^{\rm e}-x_3^{\rm p}=z$, from Eqs. (\ref{equation}) and
 (\ref{trans}) we obtain
\begin{eqnarray}\label{differential4}\hspace{-2.5cm}
\left[{\rm i}{\hat{
\partial_\|}}- \frac{\hat{P_\|}}{2}-m+{\rm i}
{\hat{\partial^{\rm e}_\perp}}-e\gamma_1A_1(x_2^{\rm
e})\right]_{\lambda\beta} \left[-{\rm i}{\hat{
\partial_\|}}-\frac{ \hat{P_\|}}{2}-m+{\rm i}
{\hat{\partial^{\rm p}_\perp}}+e\gamma_1A_1(x_2^{\rm
p})\right]_{\mu\nu}
 \nonumber\\
 %\hspace{-2cm}
 \times
\begin{tabular}{c}{\Large$
 [\eta$}$(t,z,x_\perp^{{\rm e},\rm p})${\Large ]}$_{\beta\nu}$
 \end{tabular}=-{\rm i} 8\pi\alpha D_{i
j}(t,z,x^{\rm e}_{1,2}-x^{\rm
p}_{1,2})~[\gamma_i]_{\lambda\beta}~[\gamma_j]
_{\mu\nu}\begin{tabular}{c}{\Large$[
 \eta$}$ (t,z,x_\perp^{{\rm e},{\rm
p}})${\Large ]}$_{\beta\nu}$\end{tabular}, \end{eqnarray} where
$x_\perp=(x_1,x_2)$,
$\hat{\partial}_\perp=\gamma_1\frac\partial{\partial x
_1}+\gamma_2\frac\partial{\partial x_2}$, $\hat{
\partial_\|}=\frac\partial{\partial t}\gamma_0+
\frac\partial{\partial z}\gamma_3$, and
$\hat{P_\|}=P_0\gamma_0-P_3\gamma_3$.

\subsection{Fourier-Ritus Expansion in eigenfunctions of the
 transversal motion}
   Expand the dependence of  solution of Eq.
(\ref{differential4}) on the transversal degrees of freedom into
the series  over the (complete set of) Ritus \cite{ritus} matrix
eigenfunctions  $E_h(x_2)$
\begin{eqnarray}\label{ex2}\hspace{-2.5cm}\begin{tabular}{c}[{\Large$\eta$}$(t,z,
x_\perp^{{\rm e},{\rm p}})]_{\mu\nu}$\end{tabular}= \sum_{h^{\rm
e} h^{\rm p}}~{\rm e}^{{\rm i} p_1^{\rm e} x_1^{\rm e}} [E^{\rm
e}_{h^{\rm e}}(x^{\rm e}_{2})]_\mu^{\lambda^{\rm e}}[E^{\rm
p}_{h^{\rm p}} (x^{\rm p}_{2})]_\nu^{\lambda^{\rm p}}{\rm e}^{{\rm
i} p_1^{\rm p} x_1^{\rm p}}\begin{tabular}{c}{
[{\Large$\eta$}$_{h^{\rm e} h^{\rm p}}(t,z)]_{\lambda^{\rm
e}\lambda^{\rm p}}$}\end{tabular}.\end{eqnarray} Here {\Large
$\eta$}$_{h^{\rm e} h^{\rm p}}(t,z)$ denote \textit{unknown}
functions that depend on the differences of the longitudinal
variables, while the Ritus matrix functions ${\rm e}^{{\rm i}
p_1x_1}E_{h}(x_{2})$ depend on the individual coordinates $x^{{\rm
e},{\rm p}}_{1,2}$ transversal to the field. The Ritus matrix
functions and the  unknown functions {\Large$\eta$}$_{h^{\rm e}
h^{\rm p}}(t,z)$ are labelled by two pairs $h^{{\rm e}},h^{{\rm
p}}$ of quantum numbers $h=(k,p_1,)$, each pair relating to one
out of the two particles in a magnetic field. The Landau quantum
number $k$ runs all nonnegative integers, $k=0,1,2,3...$, while
$p_1$ is the particle momentum component along the transversal
axis 1. Recall that the potential $A_\mu(x)$ (\ref{asym}) does not
depend on $x_1$, so that $p_1$ does conserve. This quantum number
is connected with the orbit center coordinate ~ $\widetilde{x_2}$~
along the axis 2 \cite{QM}, $p_1= - \widetilde{x_2}eB$.

The matrix functions ${\rm e}^{{\rm i} p_1x_1}E^{{\rm e},{\rm
p}}_{h}(x_{2})$ for transverse motion in the magnetic field
(\ref{asym}), relating in (\ref{ex2}) to electrons (e) and
positrons (p), are $4\times 4$ matrices, formed, in the spinor
representation, by four eigen-bispinors of the operator $({-\rm
i}\hat{\partial}_\perp\pm e\hat{A})^2$
\begin{eqnarray}\label{eigen2}({-\rm i}\hat{\partial}_\perp\pm
e\hat{A})_{\mu\nu}^2 {\rm e}^{{\rm i} p_1x_1}[E^{{\rm e},{\rm
p}}_h(x_2)]^{(\sigma,\gamma)}_\nu= -2eBk{\rm e}^{{\rm i}
p_1x_1}[E^{{\rm e},{\rm p}}_h(x_2)]^{(\sigma,\gamma)}_\mu,
\end{eqnarray}
placed, as columns, side by side \cite{ritus}. Here the upper and
lower signs relate to electron and positron, respectively, while
$\sigma=\pm 1$ and $\gamma=\pm 1$ are eigenvalues of the operators
\begin{eqnarray}\label{gammafive}\Sigma_3=\left(\begin{tabular}{cc}$\sigma_3$&0\\0&
$\sigma_3$\end{tabular}\right),\qquad -{\rm
i}\gamma_5=\left(\begin{tabular}{cc}-I&0\\0&I\end{tabular}
\right), \end{eqnarray} diagonal in the spinor representation, to
which the same 4-spinors are eigen-bispinors \cite{footnote2}
\begin{eqnarray}\label{sg}-{\rm i}\gamma_5E_h^{(\sigma,\gamma)}=\gamma
E_h^ {(\sigma,\gamma)},\quad\Sigma_3E_h^{(\sigma,\gamma)}=\sigma
E_h^ {(\sigma,\gamma)}. \end{eqnarray} The couple of indices
$\lambda=(\sigma, \gamma)$ is united into one index $\lambda$ in
the expansion (\ref{ex2}), $\lambda=1,2,3,4$ according to the
convention: $(+1,-1)=1,~\; (-1,-1)=2, \;(+1,+1)=3,\; (-1,+1)=4$.
With this convention, the set of 4-spinors
$~[E_h(x_{2}]^{(\sigma,\gamma)}_\mu=E_{h}(x_{2})_\mu^{\lambda}~$
can be dealt with as a $4\times 4$ matrix, the united index
$\lambda$ spanning a matrix space, where the usual algebra of
$\gamma$ -matrices may act. Correspondingly,  in (\ref{ex2}) the
unknown function {\Large$[\eta$}$_{h^{\rm e} h^{\rm
p}}(t,z)${\Large$]$}$_{\lambda^{\rm e}\lambda^{\rm p}}$ is a
matrix in the same space, and contracts with the Ritus matrix
function.

Following \cite{ritus}, the matrix functions in expansion
(\ref{ex2}) can
 be written in the block form
 as  diagonal matrices
\begin{eqnarray}\label{ritus1}%\hspace{-3cm}
{\rm e}^{{\rm i} p_1x_1} E^{{\rm e},{\rm
p}}_h(x_{2})=\left(\begin{tabular}{cc}$a^{{\rm e},{\rm
p}}(h;x_{1,2}) $&0\\0&$a^{{\rm e},{\rm
p}}(h;x_{1,2})$\end{tabular}\right),
\nonumber\\%\quad
a^{{\rm e},{\rm p}}(h;x_{1,2})=\left(\begin{tabular}{cc}$a^{{\rm
e},{\rm p}}_{+1}(h;x_{1,2})$&0\\
0&$a^{{\rm e},{\rm
p}}_{-1}(h;x_{1,2})$\end{tabular}\right).\end{eqnarray} Here
$a^{{\rm e},{\rm p}}_\sigma(h;x_{1,2})$ are eigenfunctions of the
two (for each sign $\pm$) operators $[\left( ({-\rm
i}\partial_\perp)_\lambda\pm eA_\lambda\right)^2 \mp\sigma eB]$
(we denote $(\partial_\perp)_\lambda=\partial/\partial
x_\lambda,~\lambda=1,2$), labelled by the two values
$\sigma=1,~-1$,
\begin{eqnarray}\label{eigen1}%\hspace{-3cm}
[\left( ({-\rm i}\partial_\perp)_\lambda\pm eA_\lambda\right)^2
\mp\sigma eB]a^{{\rm e},{\rm p}}_\sigma(h;x_{1,2})= 2eBk a^{{\rm
e},{\rm p}}_\sigma(h;x_{1,2}),\end{eqnarray} namely, (we omit the
subscript "1" by $p_1$ in what follows)
\begin{eqnarray}\label{asigma}a^{{\rm e},{\rm p}}_\sigma(h;x_{1,2})=
{\rm e}^{{\rm i}
px_1}U_{k+\frac{\pm\sigma-1}{2}}\left[\sqrt{eB}\left(x_2\pm\frac{p
}{eB}\right)\right],\quad k=0,1,2...,\end{eqnarray} with
\begin{eqnarray}\label{parab1}U_n(\xi)=\exp\left\{-\frac{\xi^2}{2}
\right\}(2^nn!\sqrt{\pi})^{-\frac{1}{2}}H_n(\xi)\end{eqnarray}
being the normalized Hermite functions ($H_n(\xi)$ are the Hermite
polynomials). Eqs. (\ref{eigen1}) are the same as (\ref{eigen2})
due to the relation
\begin{eqnarray}\label{sootn}({\rm i}\hat{\partial}_\perp\mp
e\hat{A})^2= -[({\rm i}\partial_\perp)_\lambda\mp eA_\lambda\,]^2
\pm eB \Sigma_3 \end{eqnarray} and to Eq. (\ref{sg}).
Simultaneously, the matrix functions (\ref{ritus1}) are
eigenfunctions to the operator $-{\rm i}\partial_1$ that commutes
with $\Sigma_3$ and $\gamma_5$ (\ref{gammafive}), and with $({\rm
i}\hat{\partial}_\perp\mp e\hat{A})_{\mu\nu}^2$. The corresponding
eigenvalue $p_1$ does not, however, appear in the r.-h. side of
(\ref{eigen1}) due to the well-known degeneracy of electron
spectrum in a constant
 magnetic field.

The orthonormality  relation for the Hermite functions
\begin{eqnarray}\label{orth} \int_{-\infty}^\infty
U_n(\xi)U_{n'}(\xi){\rm d}\xi=\delta_{nn'}.\end{eqnarray} implies
the orthogonality of the Ritus matrix eigenfunctions in the form
\begin{eqnarray}\label{orth2}\sqrt{eB}\int
E^*_{h}(x_{2})_\mu^{\lambda}E_{h^\prime}(x_{2})_\mu^{\lambda^\prime}{\rm
d} x_{2}= \delta_{kk^\prime}\delta_{\lambda\lambda^\prime}.
\end{eqnarray} As a matter of fact, the matrix functions
$E_h(x_2)$ are real, and we henceforth omit the complex
conjugation sign "$^*$".

The matrix functions ${\rm e}^{{\rm i} px_1}E^{{\rm e},{\rm
p}}_h(x_{2})$
 (\ref{ritus1}) commute with the longitudinal part\\ $~\pm{\rm i}\hat
 {\partial}_\|-\hat{P}_\|/2-m$~ of the Dirac
 operator in (\ref{differential4}), owing to the
 commutativity
property\begin{eqnarray}\label{comm}[E_h(x_2),\gamma_{0,3}]_-=0,
 \end{eqnarray} and
are \cite{ritus}, in a sense, matrix eigenfunctions of the
transversal part of the Dirac operator (not only of its square
(\ref{eigen2}))
\begin{eqnarray}\label{vazhnoe}({\rm i}\hat{\partial}_\perp \mp
e\hat{A}){\rm e}^{{\rm i} px_1}E^{{\rm e},{\rm p}}_h(x_{2})=
\pm\sqrt{2eBk}\;{\rm e}^{{\rm i} px_1}E^{{\rm e},{\rm
p}}_h(x_{2})\gamma_1.
\end{eqnarray} The Landau quantum number $k$ appears here as a
"universal eigenvalue" thanks to the mechanism, easy to trace,
according to which the differential operator in the left-hand side
of Eq. (\ref{vazhnoe}) acts as a lowering or rising operator on
the functions (\ref{parab1}), whereas the matrix $\sigma_2$,
involved in $\gamma_2$, interchanges the places the functions
$U_{k},$ $U_{k-1}$ occupy in the columns. Contrary to relations,
which explicitly include the variable $\sigma$, whose value forms
the number of the corresponding column, relations (\ref{eigen2}),
(\ref{vazhnoe}), (\ref{comm}), and the first relation in
(\ref{sg}) are covariant with respect to passing to other
representation of $\gamma$-matrices, where the matrix $E_h(x_2)$
may become non-diagonal.

\subsection{Equation for the Fourier-Ritus transform of the
Bethe-Salpeter amplitude}  Now we are in a position to use
expansion (\ref{ex2}) in Eq. (\ref{differential4}). We left
multiply it by $(2\pi)^{-2}eB~{\rm e}^{-{\rm i} \overline{p}^{\rm
e} x^{\rm e}_1}E^{{\rm e}}_{\overline{h}^{{\rm e}}}(x_2^{{\rm
e}}){\rm e}^{-{\rm i} \overline{p}^{\rm p}x^{\rm p}_1}E^{\rm p
}_{\overline{h}^{\rm p }}(x_{2}^{\rm p })$, then integrate over
${\rm d}^2x^{\rm e }_{1,2}\;{\rm d}^2x^{\rm p}_{1,2}$. After using
(\ref{vazhnoe}) and (\ref{comm}), and exploiting the
orthonormality relation (\ref{orth2}) for the summation over the
quantum numbers $h^{\rm e,p}=(k^{\rm e,p}, p_1^{\rm e,p}),$ the
following expression is obtained for the left-hand side of the
Fourier-Ritus-transformed Eq. (\ref{differential4}):
\begin{eqnarray}\label{levaya}\hspace{-1cm}\left[{\rm i}{\hat{
\partial_\|}}- \frac{\hat{P_\|}}{2}-m-\gamma_1 \sqrt{2eBk^{{\rm
e}}}\right] _{\lambda\lambda^{\rm e}}\left[-{\rm i}{\hat{
\partial_\|}}- \frac{\hat{P_\|}}{2}-m+\gamma_1 \sqrt{2eBk^{\rm
p}}\right]_{\mu\lambda^{\rm p}}\begin{tabular}{c}[{\Large$
\eta$}$_{h^{{\rm e}}h^{\rm p}}(t,z)]_{\lambda^{\rm e}\lambda^{\rm
p}}$\end{tabular}.
\end{eqnarray}
We omitted the bars over the quantum numbers.

Taking the expression
\begin{eqnarray}\label{photon1}D_{ij}(t,z,x_{1,2}^{\rm e}-x_{1,2}^{\rm
p})=
  \frac{g_{ij}}{{\rm i} 4\pi^2}[t^2-z^2-(x_1^{\rm e}-x_1^{\rm
p})^2-(x_2^{\rm e}-x_2^
  {\rm p})^2]
  ^{-1}, \end{eqnarray} for the photon propagator in the Feynman gauge,
we may then
  write the
  right-hand side of Ritus-transformed
 Eq. (\ref{differential4})
 as\begin{eqnarray}\label{rhs}\hspace{-1cm}\frac\alpha{2\pi^3}
\int{\rm d} p^{\rm e}~{\rm d} p^{\rm p}~ \sum_{k^{\rm e} k^{\rm
p}}~~g_{ij}~\int [E^{{\rm e}}_{\overline{h}^{{\rm e}}}(x_{2}^{{\rm
e}}) \gamma_i E^e_{h^e}(x^e_{2})]_{\lambda\lambda^{\rm e}}\;
[E^p_{\overline{h}^p}(x^p_{2}) \gamma_jE^{\rm p }_{{h}^{\rm p
}}(x_{2}^{\rm p })]_{\mu\lambda^{\rm
p}}\begin{tabular}{c}[{\Large$ \eta$}$_{h^{{\rm e}}h^{\rm
p}}(t,z)]_{\lambda^{\rm e}\lambda^{\rm p}}$\end{tabular}\nonumber
\\
\times \frac{{\rm e}^{{\rm i} (p^{\rm e}-\overline{p}^{\rm e}
)x_1}~{\rm e}^{{\rm i} (p^{\rm p}-\overline{p}^{\rm p}
)x_1}~\;eB{\rm d}^2 x_{1,2}^e~{\rm d}^2 x_{1,2}^p} {z^2+(x_1^{\rm
e}-x_1^{\rm p})^2+(x_2^{\rm e}-x_2^{\rm p})^2-t^2}\,,
\end{eqnarray}
Integrating explicitly the exponentials in (\ref{rhs}) over the
variable $X=(x_1^{\rm e}+x_1^{\rm p})/2$, we obtain the following
expression:
\begin{eqnarray}\label{rhs2}\hspace{-2cm}
\frac\alpha{\pi^2}\int{\rm d} p~{\rm d}
P_1~\delta(\overline{P}_1-P_1) \sum_{k^ek^p}~~g_{ij}\int[E^{{\rm
e}}_{\overline{h}^{{\rm e}}}(x_{2}^{{\rm e}}) \gamma_i
E^e_{h^e}(x^e_{2})]_{\lambda\lambda^{\rm e}}\;
[E^p_{\overline{h}^p}(x^p_{2}) \gamma_jE^{\rm p }_{{h}^{\rm p
}}(x_{2}^{\rm p })]_{\mu\lambda^{\rm p}}\nonumber\\
\times
\begin{tabular}{c}[{\Large$ \eta$}$_{h^{{\rm e}}h^{\rm
p}}(t,z)]_{\lambda^{\rm e}\lambda^{\rm p}}$\end{tabular}
\frac{\exp ({\rm i} x(\overline{p}-p)){\rm d} x}
{z^2+x^2+(x_2^{\rm e}-x_2^{\rm p})^2-t^2}\;eB{\rm d} x_2^{\rm
e}{\rm d} x_2^{\rm p}, \end{eqnarray} where the new integration
variables
 $x=x_1^{\rm e}-x_1^{\rm p}$, $P_1=p^{\rm e}+p^{\rm p}$, $p=(p^{\rm
e}-p^{\rm p})/2$ and
 the new definitions
 $\overline{P}_1=\overline{p}^{\rm e}+\overline{p}^{\rm p}$,
$\overline{p}=
 (\overline{p}^{\rm e}-\overline{p}^{\rm p})/2$ have been introduced.
The pairs of quantum numbers in (\ref{rhs2})
are\begin{eqnarray}\label{pairs2} \overline{h}^{{\rm e},{\rm
p}}=(\overline{k}^{{\rm e},{\rm p}},
\frac{\overline{P}_1}{2}\pm\overline{p}),\qquad {h}^{{\rm e},{\rm
p}}=(k^{{\rm e},{\rm p}}, \frac{{P}_1}{2}\pm{p}).
\end{eqnarray}
Hence the arguments of the functions (\ref{asigma}) in
(\ref{rhs2}) are:
\begin{eqnarray}\label{arg}\hspace{-1cm}
\sqrt{eB}\left(x_2^{\rm
e}+\frac{\overline{P}_1+2\overline{p}}{2eB}\right), \,\,
\sqrt{eB}\left(x_2^{\rm e}+\frac{{P}_1+2{p}}{2eB}\right),\,\,
\sqrt{eB}\left(x_2^{\rm p}
-\frac{\overline{P}_1-2\overline{p}}{2eB}\right),\,\,
\sqrt{eB}\left(x_2^{\rm p} -\frac{{P}_1-2{p}}{2eB}\right),\quad
\end{eqnarray}
successively  as the functions $E_h(x_{1,2})$ appear in
(\ref{rhs2}) from left to right.
 After
fulfilling the integration over ${\rm d} P_1$ with the use of the
$\delta$-function, introduce the new integration variable
$q=p-\overline{p}$ instead of $p$, and the integration variables
$\overline{x}_2^{\rm e}=x_2^{\rm
e}+(\overline{P}_1+2\overline{p})/2eB$, $\overline{x}_2^{\rm
p}=x_2^{\rm p}-(\overline{P}_1-2p)/2eB$ instead of $x_2^{\rm e}$
and $x_2^{\rm p}$. Then (\ref{rhs2}) may be written as
\begin{eqnarray}\label{rhs3}\hspace{-2cm}\frac\alpha{\pi^2}\int{\rm d}
q~ \sum_{k^{\rm e} k^p}~~g_{ij}~\int [E^{{\rm
e}}_{\overline{h}^{{\rm e}}}(x_{2}^{{\rm e}}) \gamma_i
E^e_{h^e}(x^e_{2})]_{\lambda\lambda^{\rm e}}\;
[E^p_{\overline{h}^p}(x^p_{2}) \gamma_jE^{\rm p }_{{h}^{\rm p
}}(x_{2}^{\rm p })]_{\mu\lambda^{\rm
p}}\begin{tabular}{c}[{\Large$ \eta$}$_{h^{{\rm e}}h^{\rm
p}}(t,z)]_{\lambda^{\rm e}\lambda^{\rm p}}$\end{tabular}\nonumber\\
\times \int\frac{\exp (-{\rm i} xq)\;{\rm d} x \;eB\;{\rm
d}\overline{x}^e_2\;{\rm d} \overline{x}_{2}^p} {z^2+
x^2+\left(\overline{x}_2^{\rm e}-\overline{x}_2^{\rm
p}-\frac{\overline{P}_1-q} {eB} \right)^2-t^2}\,.\end{eqnarray}
Now the pairs of quantum numbers in (\ref{rhs3})
are\begin{eqnarray}\label{pairs3} \overline{h}^{{\rm e},{\rm
p}}=(\overline{k}^{{\rm e},{\rm p}},
\frac{\overline{P}_1}{2}\pm\overline{p}),\qquad {h}^{{\rm e},{\rm
p}}=(k^{{\rm e},{\rm p}}, \frac{\overline{P}_1}{2}\pm
q\pm\overline{p}).\end{eqnarray} Hence the arguments of the
functions (\ref{asigma}) in (\ref{rhs3}) from left to right are
\begin{eqnarray}\label{arg3}\hspace{-1cm}\sqrt{eB}\overline{x}^{\rm
e}_2,\quad \sqrt{eB}\left(\overline{x}^{\rm e}_2+\frac{q
}{eB}\right),\quad \sqrt{eB} \left(\overline{x}^{\rm
p}_2-\frac{q}{eB}\right),\quad \left(\sqrt{eB}\overline{x}^{\rm
p}_2\right). \end{eqnarray}

\subsection{Adiabatic approximation}
  Now we aim at passing to the large magnetic field regime
  in the Bethe-Salpeter equation, with (\ref{levaya}) as the
  left-hand side and (\ref{rhs3}) as the right-hand side.
Define the dimensionless integration  variables $w=x\sqrt{eB}$,
$q'=q/\sqrt{eB}$, $\xi^{\rm e,p}=\overline{x}^{\rm
e,p}_2\sqrt{eB}$ in function (\ref{rhs3}). Then it takes the form
\begin{eqnarray}\label{rhs4}\hspace{-1cm} \frac\alpha{\pi^2}\int{\rm d}
q'~ \sum_{k^{\rm e} k^{\rm p}}~~g_{ij}~\int [E^{{\rm
e}}_{\overline{h}^{{\rm e}}}(x_{2}^{{\rm e}}) \gamma_i E^{\rm
e}_{h^{\rm e}}(x^e_{2})]_{\lambda\lambda^{\rm e}}\; [E^{\rm
p}_{\overline{h}^{\rm p}}(x^{\rm p}_{2}) \gamma_jE^{\rm p
}_{{h}^{\rm p }}(x_{2}^{\rm p })]_{\mu\lambda^{\rm
p}}\begin{tabular}{c}[{\Large$ \eta$}$_{h^{{\rm e}}h^{\rm
p}}(t,z)]_{\lambda^{\rm e}\lambda^{\rm p}}$\end{tabular}\nonumber\\
\times \int\frac{\exp (-{\rm i} wq'){\rm d} w {\rm d}\xi^{\rm
e}{\rm d}\xi^{\rm p}} {z^2+ \frac{w^2}{eB}+\frac 1{eB}
\left(\xi^{\rm e}-\xi^{\rm p}-\frac{\overline{P}_1}{\sqrt{eB}}-q'
\right)^2-t^2}\,. \end{eqnarray} The pairs of quantum numbers in
(\ref{rhs4}) are
\begin{eqnarray}\label{pairs4}
\overline{h}^{{\rm e},{\rm p}}=(\overline{k}^{{\rm e},{\rm p}},
\frac{\overline{P}_1}{2}\pm\overline{p}),\quad {h}^{{\rm e},{\rm
p}}=(k^{{\rm e},{\rm p}}, \frac{\overline{P}_1}{2}\pm
q^\prime\sqrt{eB}\pm\overline{p}).\end{eqnarray} The arguments of
the functions (\ref{asigma}) in (\ref{rhs4}) from left to right
are
\begin{eqnarray}\label{arg4} \xi^{\rm e},\quad \xi^{\rm e}+q',\quad
\xi^{\rm p}-q',\quad \xi^{\rm p}.  \end{eqnarray}

When considering the large field behavior we admit for
completeness that the difference between the centers of orbits
along the axis 2 $\;\widetilde{x}^{\rm e}_2-\widetilde{x}^{\rm
p}_2=-\frac{\overline{P}_1}{eB}$ ~ may be kept finite, in other
words that the transversal momentum $\overline{P}_1$ grows
linearly with the field. We shall see that that big transversal
momenta do not contradict  dimensional compactification, but
produce an extra regularization of the light-cone singularity.

In the region, where the 2-interval $(z^2-t^2)^{1/2}$ essentially
exceeds the Larmour radius $L_B=(eB)^{-1/2}$,
\begin{eqnarray}\label{domain}z^2-t^2 \gg L_B^2\,,\end{eqnarray}
one may neglect the dependence on the integration variables $w$
and later on $\xi^{\rm e,p}$ in the denominator. Integration over
$w$ produces $2\pi\delta (q')$, which annihilates the dependence
on $q'$ in the arguments (\ref{arg3}) of the Hermite functions,
and they all equalize.

Let us depict this mechanism in more detail. Fulfill explicitly
the integration over ${\rm d} w$ in (\ref{rhs4}):
\begin{eqnarray}\label{explicit}
%\hspace{-2.5cm}
\int\frac{\exp (-{\rm i} wq')\;{\rm d} w} {z^2-t^2+\frac{w^2}{eB}+
\frac{A^2}{eB}}=
\frac{\sqrt{eB}\pi}{\sqrt{z^2-t^2+\frac{A^2}{eB}}}
\nonumber\\
\times\left[ \theta(q^\prime)\exp\;
[-q^\prime\sqrt{eB(z^2-t^2)+A^2}\,] + \theta(-q^\prime)\exp\;
[\;q^\prime\sqrt{eB(z^2-t^2)+A^2}\;]\right],\end{eqnarray}
where\begin{eqnarray}\label{A} A^2=
\left(\xi^e-\xi^p-\frac{\overline{P}_1}{\sqrt{eB}}-q' \right)^2
\end{eqnarray} and $\theta(q^\prime)$ is the step
function,\begin{eqnarray}\label{step}
\theta(q^\prime)=\left\{\begin{tabular}{ccc}1 & when
&$q^\prime>0,$
\\$\frac 1{2}$ & when & $q^\prime=0,$\\0&
when &$ q^\prime<0.$
\end{tabular}\right.\end{eqnarray} Due to the decreasing exponential
in (\ref{parab1}) the variables $\xi^{{\rm e},{\rm p}}$ do not
exceed unity in the order of magnitude and can be neglected as
compared to  $\frac{\overline{P}_1}{\sqrt{eB}}$ in (\ref{A}).
Unless $q^\prime$ is  large it may be neglected as compared to the
same term in (\ref{A}), too. Then
$A^2=\frac{\overline{P}_1^2}{eB}$, and after (\ref{explicit}) is
substituted in (\ref{rhs4}) and integrated over ${\rm d} q^\prime$
the contribution comes only from the integration within the
shrinking region $|q^\prime
|<(eB[z^2-t^2+\frac{\overline{P}_1^2}{(eB)^2}])^{-\frac 1{2}}$.
Then $q^\prime$ can be also neglected in the arguments
(\ref{arg4}). If, contrary to the previous assumption, we admit
that $|q^\prime |$ is of the order of
$\frac{\overline{P}_1}{\sqrt{eB}}\sim\sqrt{eB}$ we see that the
exponentials in (\ref{explicit}) fast decrease with the growth of
the magnetic field as $\exp\,[-eB(z^2-t^2)]$, and therefore such
values of $|q^\prime |$ do not contribute to the integration. If
we admit, last, that $|q^\prime |\gg
|\frac{\overline{P}_1}{\sqrt{eB}}|$, we find that the contribution
$\exp\,[-|q^\prime|\sqrt{eB(z^2-t^2)+(q^\prime)^2}\;]$ from the
integration over such values is still smaller. Thus, we have
justified the possibility to omit the dependence on $q^\prime$ in
(\ref{A}) and in (\ref{arg4}), and also on $\xi^{{\rm e},{\rm p}}$
in (\ref{A}). Now we can perform the integration over ${\rm d}
q^\prime$ to obtain the following expression for (\ref{rhs4}):
\begin{eqnarray}\label{rhs5}\hspace{-1.5cm}
\frac{2\alpha\pi^{-1}}{z^2+
\frac{{P}_1^2}{(eB)^2}-t^2}\sum_{k^{\rm e} k^{\rm p}}g_{ii}
\int[E^{{\rm e}}_{\overline{h}^{{\rm e}}}(x_{2}^{{\rm e}})
\gamma_i E^{\rm e}_{h^{\rm e}}(x^{\rm
e}_{2})]_{\lambda\lambda^{\rm e}}{\rm d} \xi^{\rm e} \int [E^{\rm
p}_{\overline{h}^{\rm p}}(x^{\rm p}_{2}) \gamma_iE^{\rm p
}_{{h}^{\rm p }}(x_{2}^{\rm p })]_{\mu\lambda^{\rm p}}{\rm
d}\xi^{\rm p}\begin{tabular}{c}[{\Large$ \eta$}$_{h^{{\rm
e}}h^{\rm p}}(t,z)]_{\lambda^{\rm e}\lambda^{\rm
p}}.$\end{tabular}\end{eqnarray} It remains yet to argue that the
limit (\ref{rhs5}) is valid also when the term
$\frac{\overline{P}_1}{eB}$ is not kept. In this case we no longer
can disregard $q^\prime$ inside $A^2$ when $q^\prime$ is less than
or of the order of unity. But we can disregard $A^2$ as compared
with $eB(z^2-t^2)$ to make sure that the integration over ${\rm d}
q^\prime$ is restricted to the  region close to zero $|q|^\prime
<[eB(z^2-t^2)]^{-1/2}$ and hence set $q^\prime=0$ in (\ref{arg4}).
The contribution of large $q^\prime$ is small as before.

 The integration over $\xi^{\rm e,p}$ of the
terms with $i=0,3$ in (\ref{rhs5}) yields the Kroneker deltas
$\delta_{k^{\rm e}\overline{k^{\rm e}}}~\delta_{k^{\rm
p}\overline{k^{\rm p}}}$ due to the orthonormality (\ref{orth}) of
the Hermite functions thanks to the commutativity (\ref{comm}) of
the Ritus matrix functions (\ref{ritus1}) with $\gamma_0$ and
$\gamma_3$. On the contrary, $\gamma_1,~\gamma_2$ do not commute
with (\ref{ritus1}). This implies the appearance of terms,
non-diagonal in Landau quantum numbers, like $\delta_{k^{\rm e},
\overline{k}^{\rm e}\pm 1}$ and $\delta_{k^{\rm p},
\overline{k}^{\rm p}\pm 1}$, in (\ref{rhs4}), proportional to
($i=1,2$):
\begin{eqnarray}\label{nondiag1}\hspace{-1cm} T^i_{\overline{k}_{\rm
e}\pm 1,\overline{k}^{\rm p}\pm 1}= \sum_{k^{\rm e} k^{\rm
p}}\int[E^{{\rm e}}_{\overline{h}^{{\rm e}}}(x_{2}^{{\rm e}})
\gamma_i E^{\rm e}_{h^e}(x^{\rm e}_{2})]_{\lambda\lambda^{\rm
e}}{\rm d}\xi^{\rm e}\;\int [E^{\rm p}_{\overline{h}^{\rm
p}}(x^{\rm p}_{2}) \gamma_iE^{\rm p }_{{h}^{\rm p }}(x_{2}^{\rm p
})]_{\mu\lambda^{\rm p}}{\rm d}\xi^{\rm
p}\begin{tabular}{c}[{\Large$ \eta$}$_{h^{{\rm e}}h^{\rm
p}}(t,z)]_{\lambda^{\rm e}\lambda^{\rm p}}$\end{tabular}
\nonumber\\\hspace{-2cm}=\sum_{k^{\rm e} k^{\rm
p}}\left(\begin{tabular}{cc}0&$-\Delta^{i}
_{\overline{k}^{\rm e} k^{\rm e}}$\\
$\Delta^{i}_{\overline{k}^{\rm e} k^{\rm
e}}$&0\end{tabular}\right)_{\lambda\lambda^{\rm e}}
\left(\begin{tabular}{cc}0&$-
\Delta^{i}_{\overline{k}^{\rm p} k^{\rm p}}$\\
$\Delta^{i}_{\overline{k}^{\rm p} k^{\rm
p}}$&0\end{tabular}\right)_{\mu\lambda^{\rm
p}}\begin{tabular}{c}[{\Large$ \eta$}$_{h^{{\rm e}}h^{\rm
p}}(t,z)]_{\lambda^{\rm e}\lambda^{\rm p}}$\end{tabular}.
\end{eqnarray}
Here $x^{{\rm e},{\rm p}}_2$ are expressed in terms of $\xi$
through the chain of the changes of variables made above starting
from (\ref{rhs}), so that all the arguments of the Hermite
functions have become equal to $\xi$. Besides,
\begin{eqnarray}\label{pairs5}
h^{{\rm e},{\rm p}}=(k^{{\rm e},{\rm p}},\overline{p}^{{\rm
e},{\rm p}}),\;\quad \overline{h}^{{\rm e},{\rm
p}}=(\overline{k}^{{\rm e},{\rm p}},\overline{p} ^{{\rm e},{\rm
p}}),\quad p^{\rm e}+p^{\rm p}=P_1.\end{eqnarray}
\begin{eqnarray}\label{Delta12}\Delta^{i}_{\overline{k}k}=\int
a^\prime(\overline{h},x_2)\;\sigma_{i}\; a^\prime(h,x_2){\rm
d}\xi,\qquad i=1,2\,,\end{eqnarray}
\begin{eqnarray}\label{delta1}\hspace{-1cm}\Delta^{(1)}_{\overline{k}k}=
\int\left(\begin{tabular}{cc}
0&$a^\prime_{+1}(\overline{h},x_2)a^\prime_{-1}(k,x_2)$\\
$a^\prime_{-1}(\overline{h},x_2)a^\prime_{+1}(h,x_2)$&0\end{tabular}
\right){\rm d}\xi=
\left(\begin{tabular}{cc}0&$\delta_{\overline{k},\;k-1}$\\
$\delta_{\overline{k},\; k+1}$&0\end{tabular}\right),
\end{eqnarray}
\begin{eqnarray}\label{delta2}\hspace{-1cm}\Delta^{(2)}_{\overline{k}k}={\rm
i} \int\left(\begin{tabular}{cc}
0&$-a^\prime_{+1}(\overline{h},x_2)a^\prime_{-1}(k,x_2)$\\
$a^\prime_{-1}(\overline{h},x_2)a^\prime_{+1}(h,x_2)$&0\end{tabular}
\right){\rm d}\xi= {\rm i}
\left(\begin{tabular}{cc}0&$-\delta_{\overline{k},\;k-1}$\\
$\delta_{\overline{k},\; k+1}$&0\end{tabular}\right).
\end{eqnarray} The prime over $a$ indicates that the exponential
$\exp ({\rm i} px_1)$ is dropped from the definitions
(\ref{ritus1}) and (\ref{asigma}). The non-diagonal Kronecker
deltas appeared, because $a^\prime_{\pm 1}(\overline{h},x_2)$ are
multiplied by $a^\prime_{\mp 1}({h},x_2)$ under the action of the
$\sigma_{1,2}$-blocks in $\gamma_{1,2}$ (\ref{gamma}). In the
final form, the matrices in (\ref{nondiag1}) are
\begin{eqnarray}\label{matr}\hspace{-2cm}
\left(\begin{tabular}{cc}0&$-\Delta^{i}
_{\overline{k} k}$\\
$\Delta^{i}_{\overline{k} k}$&0\end{tabular}\right)=\frac
1{2}\left(\gamma_1(\pm\delta_{\overline{k},k-1}+\delta_{\overline{k},k+1})+
{\rm
i}\gamma_2(\pm\delta_{\overline{k},k-1}-\delta_{\overline{k},k+1})
\right),\end{eqnarray} with the upper sign relating to $i=1$ and
the lower one to $i=2$. Now Eq. (\ref{differential4}) acquires the
following form:
\begin{eqnarray}\label{chain}\hspace{-1cm}\left[{\rm i}{\hat{
\partial_\|}}- \frac{\hat{P_\|}}2-m-\gamma_1 \sqrt{2eBk^{{\rm
e}}}\right] _{\lambda\lambda^{\rm e}}\left[-{\rm i}{\hat{
\partial_\|}}- \frac{\hat{P_\|}}2-m+\gamma_1 \sqrt{2eBk^{\rm
p}}\right]_{\mu\lambda^{\rm p}}\begin{tabular}{c}[{\Large$
\eta$}$_{h^{{\rm e}}h^{\rm p}}(t,z)]_{\lambda^{\rm e}\lambda^{\rm
p}}$\end{tabular} \nonumber\\\hspace{-1cm}=
\frac{2\alpha\pi^{-1}}{z^2+
\frac{{P}_1^2}{(eB)^2}-t^2}~\left(\sum_{i=0,3}g_{ii}[\gamma_i
]_{\lambda\lambda^{\rm e}} [\gamma_i]_{\mu\lambda^{\rm
p}}\begin{tabular}{c}[{\Large$ \eta$}$_{h^{{\rm e}}h^{\rm
p}}(t,z)]_{\lambda^{\rm e}\lambda^{\rm p}}$\end{tabular}-
\sum_{i=1,2}T^{(i)}_{{k}^{\rm e}\pm 1,\;{k}^{\rm p}\pm 1}\right).
\end{eqnarray}
The bars over quantum numbers are omitted. This equation is
degenerate with respect to the difference of the electron and
positron momentum components $p=(p^{\rm e}-p^{\rm p})/2$ across
the magnetic field, but does depend on its transversal
center-of-mass momentum $P_1=(p^{\rm e}+p^{\rm p})$. This
dependence is present, however, only for sufficiently large
transverse momenta $P_1$.

At the present step of adiabatic approximation we have come, for
high magnetic field, to the chain of Eqs. (\ref{chain}), in which
the unknown function for a given pair of Landau quantum numbers
$k^{\rm e},k^{\rm p}$ is tangled with the same function with the
Landau quantum numbers both shifted by $\pm 1$ (in contrast to the
general case of a moderate magnetic field, where these numbers may
be shifted by all positive and negative integers). To be more
precise, the chain consists of two mutually disentangled
sub-chains. The first one includes all functions with the Landau
quantum numbers $k^{\rm e},\;k^{\rm p}$ both even or both odd, and
the second includes their even-odd and odd-even combinations. We
discuss the first sub-chain since it contains the lowest function
with $k^{\rm e}=k^{\rm p}=0$. We argue now that there exists a
solution to the first sub-chain of Eqs. (\ref{chain}), for which
all {\Large$\eta$}$_{k^{\rm e} ,p_1^{\rm e};\;k^{\rm p} ,p_1^{\rm
p}}(t,z)$ disappear if at least one of the quantum numbers $k^{\rm
e},k^{\rm p}$ is different from zero. Indeed, for $k^{\rm
e}=k^{\rm p}=0$ Eqs. (\ref{chain}) then reduces to the closed set
\begin{eqnarray}\label{closed}[{\rm i}{\hat{
\partial_\|}}- \frac{\hat{P_\|}}2-m]
_{\lambda\lambda^{\rm e}}[-{\rm i}{\hat{
\partial_\|}}- \frac{\hat{P_\|}}2-m]_{\mu\lambda^{\rm p}}
\begin{tabular}{c}[{\Large$
\eta$}$_{0,p_1^{{\rm e}};0,p_1^{\rm p}}(t,z)]_{\lambda^{\rm
e}\lambda^{\rm p}}$\end{tabular}
\nonumber\\%\hspace{-3cm}
= \frac{2\alpha\pi^{-1}}{z^2+
\frac{{P}_1^2}{(eB)^2}-t^2}~\sum_{i=0,3}g_{ii}[\gamma_i]
_{\lambda\lambda^{\rm e}} [\gamma_i]_{\mu\lambda^{\rm
p}}\begin{tabular}{c}[{\Large$ \eta$}$_{0,p_1^{{\rm e}};0,p_1^{\rm
p}}(t,z)]_{\lambda^{\rm e}\lambda^{\rm p}}$\end{tabular},\quad
p_1^{\rm e}+p_1^{\rm p}=P_1.\end{eqnarray} In writing it we have
returned to the initial designation of the electron and positron
transverse momenta $p^{\rm e,\rm p}_1$. Denote for simplicity
{\Large$\eta$}$_{k^{\rm e} k^{\rm p}}=${\Large$\eta$}$_{k^{\rm
e},p_1^{\rm e}; k^{\rm p},p_1^{\rm p}}(t,z).$ If  we consider Eqs.
(\ref{chain}) with $k^{\rm e}=k^{\rm p}=1$ for
{\Large$\eta$}$_{11}$ we shall have a nonzero contribution in the
right-hand side, proportional to {\Large$\eta$}$_{00}$ coming from
$T^i_{k^{\rm e}-1,k^{{\rm p}}-1}$, since the other contributions
{\Large$\eta$}$_{11},
${\Large$\eta$}$_{22}$,{\Large$\eta$}$_{20}$,
{\Large$\eta$}$_{02}$ are vanishing according to the assumption.
As the left-hand side of Eq. (\ref{chain}) now contains a term,
infinitely growing with the magnetic field $B$, it can be only
satisfied with the function {\Large$\eta$}$_{11}$, infinitely
diminishing with $B$ in the domain (\ref{domain}) as
\begin{eqnarray}\label{eta11}
\hspace{-2cm}\begin{tabular}{c}[{\Large$
\eta$}$_{11}]_{\lambda\mu}$\end{tabular}=%\nonumber\\=
-\frac {1}{2eB}\frac{\alpha\pi^{-1}}{z^2+
\frac{{P}_1^2}{(eB)^2}-t^2}~\sum_{i=0,3}g_{ii}[\gamma_1\gamma_i]_{\lambda\lambda^{\rm
e}} [\gamma_1\gamma_i]_{\mu\lambda^{\rm
p}}\begin{tabular}{c}[{\Large$ \eta$}$_{00}]_{\lambda^{\rm
e}\lambda^{\rm p}}$\end{tabular},\end{eqnarray} in accord with the
assumption made.  Thus, the assumption that all Bethe-Salpeter
amplitudes with nonzero Landau quantum numbers are zero in the
large-field case is consistent. We state that a solution to the
closed set (\ref{closed}) for {\Large$\eta$}$ _{0,p_1^{{\rm
e}};0,p_1^{\rm p}}(t,z)$ with all the other components equal to
zero is a solution to the whole chain (\ref{chain}).

The derivation given in this Subsection realizes formally the
known heuristic argument that, for high magnetic field, the
spacing between Landau levels is very large and hence the
particles taken in the lowest Landau state remain in it.
Effectively, only the longitudinal degree of freedom survives for
large $B$, the space-time reduction taking place. Eq.
(\ref{closed}) is a fully relativistic two-dimensional set of
equations with two space-time arguments $t$ and $z$ and two
gamma-matrices $\gamma_0$ and $\gamma_3$ involved. Since, unlike
the previous works \cite{leinson1}, \cite{ShUs}, \cite{leinson2},
neither the famous equal-time Ansatz for the Bethe-Salpeter
amplitude \cite{schweber}, nor any other assumption concerning the
non-relativistic character of the internal motion inside the
positronium atom was made, the equation derived is valid for
arbitrary strong binding. It will be analyzed for the extreme
relativistic case in the next Section.

The two-dimensional equation (\ref{closed}) is valid in the
space-like domain (\ref{domain}). It is meaningful provided that
its solution is concentrated in this domain. In non-relativistic
or semi-relativistic consideration it is often accepted that the
wave function is concentrated within the Bohr radius $a_{\rm
0}=(\alpha m)^{-1}\simeq 0.5\times 10^{-8}$ cm. It is then
estimated that the corresponding analog of  asymptotic equation
(\ref{closed}) holds true when $a_{0}\gg L_B$, i.e. for the
magnetic fields larger than $B_{\rm a}= \alpha^2m^2/e\simeq
2.35\times 10^{9}$ G. This estimate, however, cannot be universal
and may be applicable at the most to the magnetic fields close to
the lower bound  where the value of the Bohr radius can be
borrowed from the theory without the magnetic field. Generally,
the question, where the wave function is concentrated, should be
answered \textit {a posteriori} by inspecting a solution to Eq.
(\ref{closed}). Therefore, one can state, how large the fields
should be in order that the asymptotic equation (\ref{closed})
might be trusted, no sooner that its solution is investigated. We
shall return to this point when we deal with the
ultra-relativistic situation.

Remind that the transverse total momentum component of the
positronium system is connected with the separation between the
centers of orbits of the electron and positron
$\;P_1/(eB)\;=~\widetilde{x}_2^e-\widetilde{x}_2^p~$ in the
transversal plane, so that the "potential" factor in Eq.
(\ref{closed}) may be expressed in the following interesting
form\begin{eqnarray}\label{interest}\frac{\alpha} {(x_0^{\rm
e}-x_0^{\rm p})^2-(x_3^{\rm e}-x_3^{\rm
p})^2-(\widetilde{x}_2^{\rm e}- \widetilde{x}_2^{\rm
p})^2},\end{eqnarray}(\textit{cf} the corresponding form of the
Coulomb potential in the semi-relativistic treatment of the
Bethe-Salpeter equation in \cite{ShUs,leinson2}- the difference
between the potentials in \cite{ShUs,leinson2} lies within the
accuracy of the adiabatic approximation). The appearance of
$P_1^2$ in the potential determines the energy spectrum dependence
upon the momentum of motion of the two-particle system across the
magnetic field like in \cite{ShUs}, \cite{leinson2}, \cite{lai}.

We shall need Eq. (\ref{closed}) in a more convenient form. First,
transcribe it as\begin{eqnarray}\label{closed2}({\rm
i}\overrightarrow{\hat{
\partial_\|}}- \frac{\hat{P_\|}}2-m)
\begin{tabular}{c}{\Large$
\eta$}$_{0,p_1^{{\rm e}};0,p_1^{\rm p}}(t,z)$\end{tabular}(-{\rm
i}\overleftarrow{\hat{
\partial_\|}}- \frac{\hat{P_\|}}2-m)^{\rm T}\nonumber\\=
\frac{2\alpha\pi^{-1}}{z^2+
\frac{{P}_1^2}{(eB)^2}-t^2}~\sum_{i=0,3}g_{ii}\gamma_i
\begin{tabular}{c}{\Large$
\eta$}$_{0,p_1^{{\rm e}};0,p_1^{\rm
p}}(t,z)$\end{tabular}\gamma_i^{\rm T}\,.
\end{eqnarray}
Here the superscript T denotes the transposition. With the help of
the relation $\gamma_i^{\rm T}=-C^{-1}\gamma_iC$, with $C$ being
the charge conjugation matrix, $C^2=1$, and the anti-commutation
relation $[\gamma_i,\gamma_5]_+=0$, $\gamma_5^2=-1$, we may write
for a new Bethe-Salpeter amplitude $\Theta(t,z)$, defined
as\begin{eqnarray}\label{theta2} \Theta(t,z)=
\begin{tabular}{c}{\Large$
\eta$}$_{0,p_1^{{\rm e}};0,p_1^{\rm
p}}(t,z)$\end{tabular}C\gamma_5,\quad \end{eqnarray} the equation
\begin{eqnarray}\label{closed3}({\rm i}\overrightarrow{\hat{
\partial_\|}}- \frac{\hat{P_\|}}2-m)
\Theta(t,z)(-{\rm i}\overleftarrow{\hat{
\partial_\|}}- \frac{\hat{P_\|}}2-m)\nonumber\\=
\frac{2\alpha\pi^{-1}}{z^2+
\frac{{P}_1^2}{(eB)^2}-t^2}~\sum_{i=0,3}g_{ii}\gamma_i
\Theta(t,z)\gamma_i \,.\end{eqnarray} The unknown function
$\Theta$ here is a 4$\times$4 matrix, which contains as a matter
of fact only four independent components. In order to
correspondingly reduce the number of equations in the set
(\ref{closed3}), one should note that the $\gamma$-matrix algebra
in two-dimensional space-time should have only four basic
elements. In accordance with this fact, only the matrices
$\gamma_{0,3}$  are involved  in (\ref{closed3}). Together with
the matrix $\gamma_0\gamma_3$ and the unit matrix $I$ they form
the basis, since $\gamma_{0,3}\cdot\gamma_0\gamma_3=\gamma_{3,0}$,
$\gamma_0^2=-\gamma_3^2=(\gamma_0\gamma_3)^2=1$,
$[\gamma_0,\gamma_3]_+= [\gamma_{0,3},\gamma_0\gamma_3]_+=0$.
Using this algebra and the general representation for the solution
\begin{eqnarray}\label{repr}\Theta=aI+b\gamma_0+c\gamma_3+d\gamma_0\gamma_3,\quad
\end{eqnarray}  one readily obtains a closed set of four first-order
differential equations for the four functions $a, b,c,d$ of $t$
and $z$. The same set will be obtained, if one replaces in Eqs.
(\ref{closed3}) and (\ref{repr}) the 4$\times$4 matrices by the
Pauli matrices (\ref{pauli}), subject to the same algebraic
relations, according, for instance, to the rule:
$\gamma_0\Rightarrow\sigma_3,\;\gamma_3\Rightarrow {\rm
i}\sigma_2,\;\gamma_0\gamma_3\Rightarrow\sigma_1$. Then Eq.
(\ref{closed}) becomes a matrix equation
\begin{eqnarray}\label{closed4}\hspace{-1cm}({\rm
i}\overrightarrow{\partial_t}\sigma_3+
\overrightarrow{\partial_z}\sigma_2
-\frac{P_0}2\sigma_3+\frac{P_3}2{\rm i}\sigma_2-m)
\vartheta(t,z)(-{\rm i}\overleftarrow{\partial_t}\sigma_3-
\overleftarrow{\partial_z}\sigma_2
-\frac{P_0}2\sigma_3+\frac{P_3}2{\rm i}\sigma_2-m)\nonumber\\=
\frac {2\alpha\pi^{-1}}{z^2+ \frac{{P}_1^2-t^2}{(eB)^2}}
~[\sigma_3 \vartheta(t,z)\sigma_3 +\sigma_2
\vartheta(t,z)\sigma_2]
\end{eqnarray}
for a 2$\times$2 matrix $\vartheta$,
\begin{eqnarray}\label{repr2}\vartheta=aI+b\sigma_3+{\rm i}
c\sigma_2+d\sigma_1.\quad \end{eqnarray} Here $I$ is the
2$\times$2 unit matrix, and functions $a, b, c, d$ are the same as
in (\ref{repr}).

\subsection{Including an external electric field}
Let us generalize the two-dimensional Bethe-Salpeter equation
obtained in the presence of a strong magnetic field by including
an external electric field, parallel to it, that is not supposed
to be strong, $E\ll B$. To this end we supplement the potential
(\ref{asym}) in Eq. (\ref{equation}) by two more nonzero
components\begin{eqnarray}\label{electric}A_{0}(x_0,x_3),
A_3(x_0,x_3))\neq 0,\quad \quad \end{eqnarray} that carry the
electric field - not necessarily constant - directed along the
axis 3. We shall use the collective notations $A_\parallel=(A_{0},
A_3)$, $x_\|=(x_0,x_3)$, $\hat{\partial}^{{\rm e},{\rm p}}_\|=
{\partial}^{{\rm e},{\rm p}}_0\gamma_0-{\partial}^{{\rm e},{\rm
p}}_3\gamma_3$, $\hat{A}_\|=A_0\gamma_0-A_3\gamma_3.$ We shall not
exploit now a representation like (\ref{trans}), but deal directly
with the Bethe-Salpeter amplitude {\Large$\chi$}$(x^{\rm e},x^{\rm
p})$ as a function of the electron and positron coordinates, and
with its Fourier-Ritus transform {\Large$\chi$}$_{h^{\rm e} h^{\rm
p}} (x^{\rm e}_\|,x^{\rm p}_\|)$ connected with {\Large$\chi$}$
(x_\|^{{\rm e}},x_\|^{{\rm p}};x_\perp^{{\rm e}},x_\perp^{{\rm
p}})$ in the same way as (\ref{ex2}). In place of Eq.
(\ref{differential4}) one should write
\begin{eqnarray}\label{differentialE}\hspace{-1cm} \left[{\rm i}{\hat{
\partial_\|}^{\rm e}}- e\hat{A_\|}(x_\|^{\rm e})-m+{\rm i}
{\hat{\partial^{\rm e}_\perp}}-e\gamma_1A_1(x_2^{\rm
e})\right]_{\lambda\beta} \left[{\rm i}{\hat{
\partial_\|}^{\rm p}}+e\hat{A}_\|(x^{\rm p}_\|)-m+{\rm i}
{\hat{\partial^{\rm p}_\perp}}+e\gamma_1A_1(x^{\rm
p}_2)\right]_{\mu\nu}
 \nonumber\\\hspace{-2cm}\times
\begin{tabular}{c}{\Large$
 [\chi$}$(x^{{\rm e},{\rm p}}_\|,x_\perp^{{\rm e},\rm p})${\Large
]}$_{\beta\nu}$
 \end{tabular}=-{\rm i} 8\pi\alpha D_{i
j}(t,z,x^{\rm e}_{1,2}-x^{\rm
p}_{1,2})~[\gamma_i]_{\lambda\beta}~[\gamma_j]
_{\mu\nu}\begin{tabular}{c}{\Large$[
 \chi$}$(x^{{\rm e},{\rm p}}_\|,x_\perp^{{\rm e},\rm p})
${\Large ]}$_{\beta\nu}$\end{tabular}\,.
\end{eqnarray}
 Thanks to the
commutativity (\ref{comm}) the rest of the procedure of the
previous Subsection remains essentially the same, and we come, in
place of (\ref{closed}), to the following two-dimensional
equation:
\begin{eqnarray}\label{closedE}\hspace{-2.5cm} \left[{\rm i}{\hat{
\partial_\|}^{\rm e}}- e\hat{A_\|}(x_\|^{\rm
e})-m\right]_{\lambda\beta} \left[{\rm i}{\hat{
\partial_\|}^{\rm p}}+e\hat{A}_\|(x^{\rm p}_\|)-m\right]_{\mu\nu}
\begin{tabular}{c}
 {\Large$[
 \chi$}$_{0,p_1^{\rm e};0,p_1^{\rm p}}(x^{\rm e}_\|, x^{\rm p}_\|)
${\Large
]}$_{\beta\nu}$\end{tabular}\nonumber\\\hspace{-2cm}=\frac{2\alpha\pi^{-1}}{z^2+
\frac{{P}_1^2}{(eB)^2}-t^2}~\sum_{i=0,3}g_{ii}~[\gamma_i]_{\lambda\beta}~[\gamma_i]
_{\mu\nu}\begin{tabular}{c}{\Large$[
 \chi$}$_{0,p_1^{\rm e};0,p_1^{\rm p}}(x^{\rm e}_\|, x^{\rm p}_\|)
${\Large ]}$_{\beta\nu}$\end{tabular}, \end{eqnarray} for a
positronium atom in a strong magnetic field placed in a moderate
electric field, parallel to the magnetic one. In order to apply
this equation to a system of two different oppositely charged
particles interacting with each other through the photon exchange
and placed into the combination of a strong magnetic and an
electric field in the same direction, say a relativistic hydrogen
atom, one should only distinguish the two masses in the first and
second square brackets in the left-hand side.

\section{Ultra-relativistic regime in a magnetic field}
In the ultra-relativistic limit, where the positronium mass is
completely compensated by the mass defect, $P_0=0$, for the
positronium at rest along the direction of the magnetic field
$P_3=0$, the most general relativistic-covariant form of the
solution (\ref{repr})
is\begin{eqnarray}\label{repr3}\Theta=I\Phi+\hat{\partial}_\|\Phi_2+\gamma_0
\gamma_3\Phi_3.\quad \end{eqnarray} The point is that
$\gamma_0\gamma_3$ is invariant under the Lorentz rotations in the
plane ($t,z$). Substituting this into (\ref{closed3}) with
$P_0=P_3=0$ we get a separate equation for the singlet component
of (\ref{repr3})
 \begin{eqnarray}\label{2D}\left( -\Box_2 + m^2\right)
\Phi(t,z)= \frac{
4\alpha\pi^{-1}\Phi(t,z)}{z^2+\frac{P_1^2}{(eB)^2}-t^2}\quad
 \quad \end{eqnarray}  and the set of equations
\begin{eqnarray}\label{other}
 \left( \Box_2+ m^2\right)
\Phi_3(t,z)= -\frac{
4\alpha\pi^{-1}\Phi_3(t,z)}{z^2+\frac{P_1^2}{(eB)^2}-t^2},\nonumber\\
(-\Box_2+m^2)\partial_t\Phi_2+2m{\rm i}\partial_z\Phi_3=0,\nonumber\\
(-\Box_2+m^2)\partial_z\Phi_2+2m{\rm i}\partial_t\Phi_3=0\quad
 \quad \end{eqnarray} for the other two components. Here $\Box_2=
 -\partial^2/\partial t^2+\partial^2/\partial z^2$ is the Laplace
  operator in two dimensions. Note the "tachyonic" sign in front of
  it in the first equation (\ref{other}).

  Let us differentiate the second equation in (\ref{other}) over $z$
   and the third one over $t$ and subtract the results from each
other.
    In this way we get that $\Box_2\Phi_3=0$. This, however,
contradicts
 the first equation in (\ref{other}) if $\Phi_3\neq 0$.
 Therefore, only $\Phi_3= 0$ is possible.
 Then, the two second equations in (\ref{other}) are satisfied,
 provided that $(-\Box_2+m^2)\Phi_2=0$. We shall concentrate in
 Eq. (\ref{2D}) in what follows.

 The longitudinal momentum along
$x_1$, or the distance between the orbit centers along $x_2$,
plays the role of the effective photon mass and a singular
potential regularizator in Eq. (\ref{2D}). The lowest state
corresponds to the zero value of the transverse total momentum
$P_1=0$. In this case Eq. (\ref{2D}) for the Ritus transform of
the Bethe-Salpeter amplitude finally becomes
\begin{eqnarray}\label{2Df}\left( -\Box_2+ m^2\right) \Phi(t,z)=
\frac{ 4\alpha\Phi(t,z)}{\pi(z^2-t^2)}.\quad \quad \quad \quad
\quad
\end{eqnarray}
We consider now the consequences of the fall-down onto the center
phenomenon present in Eq. (\ref{2Df}), formally valid for an
infinite magnetic field, and the alterations introduced by its
finiteness.

\subsection{Fall-down onto the center in the Bethe-Salpeter amplitude
for strong magnetic field} In the most symmetrical case, when the
wave function $\Phi (x)=\Phi(s)$ does not depend on the hyperbolic
angle $\phi$ in the space-like region of the two-dimensional
Minkowsky space, $t=s\sinh\phi,\;
z=s\cosh\phi,\;s=\sqrt{z^2-t^2}$, Eq. (\ref{2Df}) becomes the
Bessel differential equation\begin{eqnarray}\label{last2}
-\frac{{\rm d}^2\Phi}{{\rm d} s^2}-\frac 1{s}\frac{{\rm
d}\Phi}{{\rm d} s}+m^2\Phi=\frac{4\alpha}{\pi s^2}\Phi.\quad
\end{eqnarray} It follows from the derivation procedure in the
previous Section II that this equation is valid within the
interval\begin{eqnarray}\label{interval}\frac 1{\sqrt{eB}}\ll
s_0\leq s\leq\infty,\quad\end{eqnarray} where the lower bound
$s_0$ depends on the external magnetic field - it should be larger
than the Larmour radius $L_B=(eB)^{-1/2}$ and tend to zero
together with it, as the magnetic field tends to infinity. The
stronger the field, the ampler the interval of validity, the
closer to the origin $s=0$ the interval of validity of this
equation extends. If the magnetic field is not sufficiently
strong, the lower bound $s_0$ falls beyond the region where the
solution is mostly concentrated and the limiting form of the
Bethe-Salpeter equation becomes noneffective, since it only
relates to the asymptotic (large $s$) region, while the rest of
the $s$-axis is served by more complicated initial Bethe-Salpeter
equation, not reducible to the two-dimensional form there. This is
how the strength of the magnetic field participates - note, that
the coefficients of Eq. (\ref{last2}) do not contain it.

Solutions of (\ref{last2}) behave near the singular point $s=0$
like $s^{\sigma}$, where \begin{eqnarray}\label{sigma1}
\sigma=\pm2\sqrt{-\frac\alpha{\pi}}.\end{eqnarray} The fall-down
onto the center \cite{QM} occurs, if $\alpha>\alpha_{\rm cr}= 0$,
\textit{i.e.,} for arbitrary small attraction, the genuine value
$\alpha =1/137$ included. This differs crucially from the case of
zero magnetic field where $\alpha_{\rm cr}=\pi/8$
\cite{newfootnote}. This difference is a purely geometrical
consequence \cite{SU05} of the dimensional reduction of the
Minkowsky space from (1,3) to (1,1).

In discussing the physical consequences of the falling to the
center we appeal to the approach recently developed by one of the
present authors as applied to the Schr$\ddot{\rm o}$dinger
equation with singular potential \cite{shabad} and to the Dirac
equation in supercritical Coulomb field \cite{shabad2}. Within
this approach the singular center looks like a
 black hole. The solutions of the differential
equation that oscillate near the singularity point are treated as
free particles emitted and absorbed by the singularity. This
treatment becomes natural after the differential equation is
written as the generalized eigenvalue problem with respect to the
coupling constant. Its solutions make a (rigged) Hilbert space and
are subject to orthonormality relations with a singular measure.
This singularity makes it possible for the oscillating solutions
to be
 normalized  to  $\delta$-functions, as free particle
wave-functions should be. The nontrivial, singular  measure that
appears in the definition of the scalar product of quantum states
in the Hilbert space of quantum mechanics introduces the geometry
of a black hole of non-gravitational origin and the idea of
horizon. The deviation from the standard quantum theory manifests
itself in this approach only when particles are so close to one
another that the mutual Coulomb field they are subjected to falls
beyond the range, where the standard theory may be referred to as
firmly established \cite{shabad2}.

Following this theory we shall be using $s_0$ as the lower edge of
the normalization box \cite{shabad,shabad2}. For doing this it is
necessary that $s_0$ be much smaller than the electron Compton
length, $s_0\ll m^{-1}\simeq 3.9\times 10^{-11}$ cm, the only
dimensional parameter in Eq. (\ref{last2}). In this case the
asymptotic regime of small distances is achieved and nothing in
the region $s<s_0$ beyond  the normalization volume - where the
two-dimensional equations (\ref{closed}), (\ref{closed3}),
(\ref{2D}), (\ref{2Df}) and hence (\ref{last2}) are not valid and
the space-time for charged particles remains four-dimensional -
may affect the problem, because  this is left behind the event
horizon.

In alternative to this, we might treat $s_0$ as the cut-off
parameter. In this case we have had to extend Eq. (\ref{last2})
continuously to the region $0\leq s\leq s_0$, simultaneously
replacing the singularity $s^{-2}$ in it by a model function of
$s$, nonsingular in the origin, say, a constant $s_0^{-2}$. In
this approach the results are dependent on the choice of the model
function which is intended to substitute for the lack of a
treatable equation in that region. Besides, the limit
$s_0\rightarrow 0$ does not exist. The latter fact implies that
the approach should become invalid for sufficiently small $s_0$,
i.e., large $B$. We, nevertheless, shall also test the
consequences of this approach later in this section to make sure
that in our special problem the result is not affected any
essentially.

\subsection{Ultimate magnetic field}

With the substitution ~~$\Phi (s)=\Psi (s) /\sqrt{s}$~~ Eq.
(\ref{last2}) acquires the standard form of a Schr$\ddot{\rm
o}$dinger equation
\begin{eqnarray}\label{last3}
-\frac{{\rm d}^2\Psi(s)}{{\rm d}
s^2}+\frac{-4\frac\alpha{\pi}-\frac
1{4}}{s^2}\Psi(s)+m^2\Psi(s)=0\,.
\end{eqnarray}
Equation (\ref{last3}) is valid in the interval
\begin{eqnarray}
 s_0\leq s\leq
 \infty,\qquad s_0\gg L_B=(eB)^{-1/2}
\end{eqnarray}
 Treating the applicability boundary $s_0$ of this equation as
the lower edge of the normalization box, as  discussed above,
$s_0\ll m^{-1}$,  we impose the standing wave boundary
condition\begin{eqnarray}\label{stand} \Psi (s_0)=0 \,,
\end{eqnarray}
 on the solution of (\ref{last3})
\begin{eqnarray}\label{mcdonald2}\Psi(s)=\sqrt{s}\;\mathcal{K}_\nu(ms),\quad
\nu={\rm i} 2\sqrt{{\alpha}/{\pi}} \simeq 0.096\,{\rm i}\quad
\end{eqnarray} that decreases at infinity. It behaves near the
singular point $s=0$ as \begin{eqnarray}\label{behave} \left(\frac
s{2}\right)^{1+ \nu}\frac 1{\Gamma(1+\nu)}-\left(\frac
s{2}\right)^{1- \nu}\frac 1{\Gamma(1-\nu)}.\end{eqnarray} Here the
Euler $\Gamma$-functions appear. Starting with a certain small
value of the argument $ms$, the McDonald function with  imaginary
index $\mathcal{K}_\nu(ms)$ (\ref{mcdonald2}) oscillates, as
$s\rightarrow 0$, passing  the zero value infinitely many times.
Therefore, if $s_0$ is sufficiently small the standing wave
boundary condition (\ref{stand}),prescribed by the theory of Refs.
\cite{shabad,shabad2}, can be definitely satisfied. Keeping to the
genuine value of the coupling constant $\alpha =1/137\,\,
(\nu=0.096~{\rm i}$) one may ask: what is the largest possible
value $s_0^{\rm max}$ of $s_0$, for which the boundary problem
(\ref{last3}), (\ref{stand}) can be solved? By demanding, in
accord with the validity condition (\ref{interval}) of Eqs.
(\ref{last2}) and (\ref{last3}), that the value of $s_0^{\rm max}$
should exceed the Larmour radius:
\begin{eqnarray}\label{B} s_0^{\rm max}\gg
(eB)^{-1/2}\quad {\rm or}\quad B\gg\frac 1{e\;(s_0^{\rm
max})^2}\,,\qquad \qquad \end{eqnarray} one establishes, how large
the magnetic field should be in order that the boundary problem
might have a solution, in other words, that the point $P_0={\bf
P}=0$ might belong to the spectrum of bound states of the
Bethe-Salpeter equation  in its initial form (\ref{equation}).

  One can use the asymptotic form of the McDonald function near zero
to see that the boundary condition (\ref{stand}) is satisfied
provided that \begin{eqnarray}\label{spectre}
\left(\frac{ms_0}{2}\right)^{2\nu}=
\frac{\Gamma(1+\nu)}{\Gamma^*(1-\nu)} \qquad  \end{eqnarray} or
\begin{eqnarray}\label{spectre2}\nu\ln\frac {ms_0}{2}={\rm
i}\arg\Gamma(\nu+1)- {\rm i} \pi n,\qquad n=0,\pm 1,\pm
2,...\qquad \end{eqnarray} Since $|\nu|$ is small we may exploit
the approximation $\Gamma(1+\nu)\simeq 1-\nu C_{\rm E},$ where
$C_{\rm E}=0.577$ is the Euler constant, to
get\begin{eqnarray}\label{spectre3}\ln\left(\frac{ms_0}{2}\right)=-\frac
n{2} {\sqrt{\frac{\pi^3}{\alpha}}}-C_{\rm E}, \quad n=1,2,...\quad
\end{eqnarray}
We have expelled the non-positive integers $n$ from here, since
they would lead to the roots for $ms_0$ of the order of or larger
than unity in contradiction to the adopted condition $s_0\ll
m^{-1}$. For such values eq. (\ref{behave}) is not valid. It may
be checked that there are no other zeros of McDonald function,
besides  (\ref{spectre3}). The maximum value for $s_0$ is provided
by $n=1$. We finally get
\begin{eqnarray}\label{spectre4}\ln\left(\frac{ms_0^{\rm
max}}{2}\right)=-\frac 1{2} {\sqrt{\frac{\pi^3}{\alpha}}}-C_{\rm
E} \nonumber\\\hspace{-2.5cm}{\rm or}\nonumber\\s_0^{\rm
max}=\frac 2{m}\exp\left\{- \frac 1{2}
{\sqrt{\frac{\pi^3}{\alpha}}}-C_{\rm E}\right\} \simeq
10^{-14}\frac 1{m}.\quad \end{eqnarray} This is fourteen orders of
magnitude smaller than the Compton length $m^{-1}=3.9\times
10^{-11}$cm and makes about $10^{-25}$ cm. Now, in accord with
(\ref{B}), if the magnetic field exceeds the ultimate
value of \begin{eqnarray}\label{final} B_{\rm
ult}=\frac{m^2}{4e}\exp\left\{\frac{\pi^{3/2}}
{\sqrt{\alpha}}+2C_{\rm E}\right\}\simeq 1.6 \times
10^{28}~B_0,\quad \end{eqnarray}
 the positronium ground state
with the center-of-mass 4-momentum equal to zero  appears. Here
$B_0=m^2/e\simeq 1.22\times 10^{13}$ Heaviside-Lorentz units is
the Schwinger critical field, or $B_0=m^2c^3/e\hbar \simeq
4.4\times 10^{13}$ G. The value of $B_{\rm ult}$ is $\sim 10^{42}$~G 
that is a few orders of
magnitude smaller than the highest magnetic field in the
vicinity of superconductive cosmic strings \cite{W85}.
Excited positronium states may also reach
the spectral point $P_\mu=0$, but this occurs for magnetic fields,
tens orders of magnitude larger than (\ref{final}) - to be found
in the same way from (\ref{spectre3}) with $n=2,3...$

The ultra-relativistic state $P_\mu=0$ has the internal structure
of what was called a \textit{confined} state in
\cite{shabad,shabad2}, i.e. the one whose wave function behaves as
a standing wave combination of free particles incoming from behind
the lower edge of the normalization box and then totally reflected
back to this edge. It decreases as $\exp (-ms)$ at large distances
like the wave function of a bound state. The effective "Bohr
radius", i.e. the value of $s$ that provides the maximum to the
wave function (\ref{mcdonald2}) makes $s_{\rm max}=0.17 m^{-1}$
(this fact is established by numerical analysis). This is
certainly much less than the standard Bohr radius $a_0=(\alpha
m)^{-1}$. Taken at the level of 1/2 of its maximum value, the
wave-function is concentrated within the limits $0.006
~m^{-1}<s<1.1~ m^{-1}$. But the effective region occupied by the
confined state is still much closer to $s=0$. The point is that
the probability density of the confined state is the wave function
squared \textit{weighted with the measure} $s^{-2}{\rm d} s$
\textit{singular in the origin} \cite{shabad}, \cite{shabad2} and
is hence concentrated near the edge of the normalization box
$s_0\simeq 10^{-25}$ cm, and not in the vicinity of the maximum of
the wave function. The electric fields at such distances are about
$10^{43}$ Volt/cm. Certainly, there is no evidence that the
standard quantum theory should be valid under such conditions.
This remark gives the freedom of applying the theory presented in
Refs. \cite{shabad}, \cite{shabad2}.

 A relation like (\ref{final}) between a Fermion mass
and the magnetic field is present in \cite{miransky}. There,
however, a different problem is studied and, correspondingly, a
different meaning is attributed to that relation: it expresses the
mass acquired dynamically by a primarily massless Fermion in terms
of the magnetic field applied to it. The  mass generation is
described by the homogeneous Bethe-Salpeter equation, whose
solution is understood \cite{fomin,miransky} as the wave function
of the Goldstone boson corresponding to the spontaneous breaking
of the chiral symmetry characteristic of the massless QED. It is
claimed, moreover, that the resulting relation between the
magnetic field and the acquired mass is independent of the choice
of the gauge for the photon propagator. The equations of Ref.
\cite{miransky} may well be read off, formally, as serving our
problem of the compensation of the positronium rest mass by the
mass defect in a magnetic field, too, and the resulting expression
may be used for determining the corresponding magnetic field,
provided that the electron mass $m$ is substituted for the
acquired mass $m_{\rm dyn}$ of \cite{miransky}. There is, however,
an important discrepancy in numerical coefficients in the
characteristic exponential between (\ref{final}) and the
corresponding formula in \cite{miransky}: the latter contains
$\exp \{\pi^{3/2}/(2\alpha)^{1/2}\}$ in place of $\exp
\{\pi^{3/2}/\alpha^{1/2}+2C_{\rm E}\}$ in (\ref{final}) and its
direct use  would lead to a more favorable estimate of the
ultimate value of the magnetic field, $2.6\times 10^{19}B_0$,
than (\ref{final}). Although the basic mechanisms, the dimensional
reduction and falling to the center, acting here and in
\cite{miransky}, are essentially the same, the procedures  are
very much different, and the origin of the discrepancy remains
unclear.
  Later, in \cite{gusynin2} the authors revised
their relation in favor of a different approximation. Supposedly,
the revised relation may be also of use in the problem of ultimate
magnetic field dealt with here.

It is interesting to compare the  value (\ref{final}) with the
analogous value, obtained earlier by the present authors (see
p.393 of Ref. \cite{ShUs}) by extrapolating the nonrelativistic
result concerning the positronium binding energy in a magnetic
field to extreme relativistic region:
\begin{eqnarray}\label{old} \left.B_{\rm
ult}\right|_{\rm
NON-REL}=\frac{\alpha^2m^2}{e}\exp\left\{\frac{2\sqrt{2}}
{\alpha}\right\} \simeq 10^{164}B_0\,. \end{eqnarray} Such is the
magnetic field that makes the binding energy of the lowest energy
state equal to ~~(-2m). (This is worth comparing with the magnetic
field, estimated \cite{duncan} as $\alpha^2\exp (2/\alpha) B_0$,
that makes the mass defect of the nonrelativistic hydrogen atom
comparable with the electron rest mass). %, estimated to be% provides
%the reaching by the electron binding energy in the nonrelativistic
%hydrogen atom the value (-m) that is estimated \cite{duncan}
%$\alpha^2\exp (2/\alpha) B_0$).
We see that the relativistically enhanced attraction has resulted
in a drastically lower value of the ultimate magnetic field.
Note the difference in the character of the essential
nonanalyticity with respect to the coupling constant: it is $\exp
(\pi\sqrt{\pi}/\sqrt{\alpha})$ in (\ref{final}) and $\exp
(2\sqrt{2}/{\alpha})$ in (\ref{old}). Another effect of
relativistic enhancement is that within the semi-relativistic
treatment of the Bethe-Salpeter equation
\cite{leinson1}-\cite{leinson2}, as well as within the one using
the Schr$\rm \ddot{o}$dinger equation \cite{loudon}, only the
lowest level could acquire unlimited negative energy with the
growth of the magnetic field, whereas according to
(\ref{spectre3}) in our fully relativistic treatment all excited
levels with $n>1$ are subjected to the falling to the center and
can reach in turn the point $P_\|=0$.

Let us see now, how the result (\ref{final}) is altered if the
cut-off procedure of Ref. \cite{QM} is used. Consider Eq.
(\ref{last3}) in the domain $s_0<s<\infty$, but replace it with
another equation\begin{eqnarray}\label{psi0}-\frac{{\rm
d}^2\Psi_0(s)}{{\rm d} s^2}-\frac{\frac {4\alpha}{\pi}+\frac
1{4}}{s_0^2}\Psi_0(s)+m^2\Psi_0(s)=0\end{eqnarray} in the domain
$0<s<s_0$. The singular potential is replaced by a constant near
the origin in (\ref{psi0}). Demand, in place of (\ref{stand}),
that
$\Psi_0(0)=0,\;[\Psi_0'(s_0)/\Psi_0(s_0))=(\Psi'(s_0)/\Psi(s_0)]$.
Then, the result (\ref{final}) will be modified by the
factor\begin{eqnarray}\label{factor}\exp\left\{-\frac
2{\sqrt{\frac{4\alpha}{\pi}+\frac
1{4}}\;\cot\left(\frac{4\alpha}{\pi}+\frac 1{4}\right)-\frac
1{2}}\right\}, \end{eqnarray} which may be taken at the value
$\alpha=0$. Thus, the result (\ref{final}) is only modified by a
factor of $\exp(-4/3)\simeq 0.25$. Generally, the estimate of the
limiting magnetic field (\ref{final}) is practically nonsensitive
to the way of cut-off, in other words to any solution of the
initial equation inside the region $0<s<s_0$, where the magnetic
field does not dominate over the mutual attraction force between
the electron and positron. This fact takes place, because the term
$(\pi^{3/2}/\sqrt{\alpha})\simeq 65$, singular in $\alpha$, is
prevailing in (\ref{final}), the details of the behavior of the
wave function close to the origin $s=0$ being not essential
against its background.

\subsection{Radiative corrections}
\subsubsection{Vacuum polarization}
We should answer the question of whether the effects of vacuum
polarization in a strong magnetic field may or may not screen the
interaction between the electron and positron in such a way as to
prevent the falling to the center in the positronium atom. It is
clear {\em aposteriory} that no matter how strong the magnetic
field is, the ultraviolet singularity  dominates over its
influence in the photon propagator, if the interval sufficiently
close to the light cone is involved. Therefore, there is a
competition between the magnetic field and this characteristic
interval, which is in our problem the Larmour radius that itself
depends on the magnetic field. We have to consider the outcome of
this competition.

 To include the effect of the
vacuum polarization we should use the photon propagator in a
magnetic field, whose influence is realized via the vacuum
polarization radiative corrections, instead of its free form
(\ref{photon1}) used above. The photon propagator in a constant
and homogeneous magnetic field has the following
approximation-independent structure \cite{batalin}-
\cite{kuznetsov}\begin{eqnarray}\label{fourier} D_{ij}(x)=\frac
1{(2\pi)^4}\int \exp ({\rm i}kx) D_{ij}(k)~{\rm d}^4k,\quad
i,j=0,1,2,3,
\end{eqnarray}\begin{eqnarray}\label{photon2} D_{ij}(k)=
\sum_{a=1}^4 D_a(k)~\frac{b_i^{(a)}~
b_j^{(a)}}{(b^{(a)})^2},\nonumber\\
%D_a%^\prime}
%(k)=\frac 1{k^2-\kappa_a(k)},\quad {\rm when}\; a=1,2,3,\quad
%D_4(k)~ {\rm is~ arbitrary}.
D_a(k)=\left\{\begin{tabular}{cc}$-[k^2+\kappa_a(k)]^{-1},$&\qquad\;
$a=1,2,3$\\arbitrary, &$a$=4\end{tabular}\;.\right.
\end{eqnarray} Here $b^{(a)}$ and $\kappa^a$ are  four eigenvectors
and four eigenvalues of the polarization operator
$\Pi_{ij}$\begin{eqnarray}\label{eigen}\Pi_i^{~j}~b^{(a)}_j=\kappa_a(k)~b^{(a)}_i.
\end{eqnarray} The eigenvectors are known in the final
form:\begin{eqnarray}\label{vectors}
b^{(1)}_i=(F^2k)_ik^2-k_i(kF^2k),\quad
b_i^{(2)}=(\widetilde{F}k)_i,\quad b_i^{(3)}=(Fk)_i,\quad
b_i^{(4)}=k_i,
\end{eqnarray}where $F$, $\widetilde{F}$ and $F^2$ are the
external electromagnetic field tensor, its dual, and its tensor
squared, respectively, contracted with the photon 4-momentum $k$.
On the contrary, the eigenvalues $\kappa_{1,2,3}(k)$ are generally
unknown - subject to approximate calculations - scalar functions
of two Lorentz-invariant combinations of the momentum and the
field, which in the special frame, where the external
electromagnetic field is given by (\ref{asym}), are $k_0^2-k_3^2$
and $k_1^2+k_2^2\equiv k_\perp^2$. The eigenvalue $\kappa_4$ is
equal to zero as a trivial consequence of the transversality
$\Pi_i^{~j}k_j=0$ of the polarization operator. The eigenvectors
(\ref{vectors}) with $a=1,2,3$ are 4-potentials of the three
photon modes, while the  dispersion laws of the corresponding
electromagnetic eigenwaves are obtained by equalizing the
denominators in (\ref{photon2}) with zero. In the special frame
the eigenvectors (\ref{vectors}), up to normalizations, are
\begin{eqnarray}\label{b}
b_i^{(1)}=k^2\left(\begin{tabular}{c}0\\$k_1$\\$k_2$\\0\end{tabular}\right)_i+
k_\perp^2\left(\begin{tabular}{c}$k_0$\\$k_1$\\$k_2$\\$k_3$\end{tabular}\right)_i,\quad
b_i^{(2)}=\left(\begin{tabular}{c}$k_3$\\0\\0\\$k_0$\end{tabular}\right)_i,\quad
b_i^{(3)}=\left(\begin{tabular}{c}0\\$k_2$\\$-k_1$\\0\end{tabular}\right)_i.
\end{eqnarray}
When calculated \cite{batalin}, \cite{shabtrudy} within the
one-loop approximation of the Furry picture (i.e. using exact
Dirac propagators in the external magnetic field without radiative
corrections) these eigenvalues have the following asymptotic
behavior \cite{kratkie2}, \cite{shabtrudy}, \cite{kuznetsov},
\cite{zhetf}, \cite{heyl} (note the difference in the signs in
front of $k^2$ due to a different metric convention used here) for
large fields $eB\gg m^2$, $eB\gg |k_{3}^2-k_0^2|$\begin{eqnarray}
\kappa_1(k_0^2-k_3^2,k_\perp^2)=\frac{-\alpha k^2}
{3\pi}\left(\ln \frac{B}{B_0}-C-1.21\right), %\nonumber \\[4pt]
\label{1} \end{eqnarray} \begin{eqnarray}
\kappa_2(k_0^2-k_3^2,k_\perp^2)=\hspace{80mm}\nonumber \\[4pt]
=\frac{\alpha Bm^2(k_0^2-k_3^2)}{\pi B_0}\exp \left(-
\frac{k_\bot^2}{2m^2}\frac{B_0}{B}
\right)\int_{-1}^1\frac{(1-\eta^2)\rm d
\eta}{4m^2-(k_0^2-k_3^2)(1-\eta^2)},\nonumber\\[4pt]
\label{2}\end{eqnarray}
\begin{eqnarray} \kappa_3(k_0^2-k_3^2,k_\perp^2)=\hspace{80mm}\nonumber
\\[4pt]
=\frac{-\alpha k^2} {3\pi}\left(\ln \frac{B}{B_0}-C \right) -\frac
\alpha{3\pi}\left[0.21k_\perp^2-1.21(k_0^2-k_3^2)\right].\hspace{20mm}
%\nonumber \\
%-\frac{\alpha}{\pi}(2k_\perp^2+k_3^2-k_0^2)
%\left(\exp\left(-\frac{k_\perp^2}{2m^2}\frac{B_0}{B}\right)-1\right).\hspace{5mm}
\label{kappa3}\end{eqnarray}% Here $\alpha=1/137$ is the fine
%structure constant,
$C=0.577$ is the Euler constant. Eqs.~(\ref{1}) and (\ref{kappa3})
are accurate up to terms, decreasing with $B$ like $(B_{cr}/B)\ln
(B/B_{cr})$ and faster. Eq.~(\ref{2}) is accurate up to terms,
logarithmically growing with $B$. In $\kappa_{1,3}$ we took also
the limit $k_\perp^2\ll(B/B_{cr})m^2$, which is not the case for
$\kappa_2$, wherein the factor $~~\exp~ (-k_\perp^2B_{0}/2m^2B)~~$
is kept different from unity.  Although the components
$\kappa_{1,2,3}$ contain the growing logarithms $\alpha\ln(B/B_0)$
%in the denominators due to eqs.(\ref{1}), (\ref{kappa3}), the
%latter do not lead to the vanishing of their contributions into
%eq.(\ref{closed}) since
 the latter are yet small for the values of the
magnetic field of the order of $B_{\rm ult}$ (\ref{final}).
This is not the case for the linearly growing part of $(\ref{2})$.

Let us inspect the contributions of the photon propagator
(\ref{photon2}) into the equation that should appear in place of
(\ref{closed}). To match the diagonal form (\ref{photon1})
corresponding to the Feynman gauge we fix the gauge arbitrariness
by  choosing \begin{eqnarray}\label{D_4}
D_4(k)=-[k^2+\kappa_1(k)]^{-1}.\end{eqnarray} In the isotropic
case where no magnetic field is present all the three nontrivial
eigenvalues are the same, $\kappa_a(k)=\kappa(k)$, $a=1,2,3$.
Then, with the choice (\ref{D_4}) in (\ref{photon2}) the photon
propagator in this limit becomes
diagonal\begin{eqnarray}\label{diagonal}D_{ij}(k)=
-\frac{1}{k^2+\kappa(k)}\sum_{a=1}^4
~\frac{b_i^{(a)}~
b_j^{(a)}}{(b^{(a)})^2}=-\frac{g_{ij}}{k^2+\kappa(k)},\end{eqnarray}
since the eigenvectors (\ref{eigen}) or (\ref{b}) make an
orthogonal basis irrespective of whether the magnetic field is
present or not.

In spite of the presence \cite{skobelev}, \cite{kratkie2} of a
term, linearly growing with the field in (\ref{2}), the component
$D_2$ does contribute in the limit of high fields into the
right-hand side of an equation to replace (\ref{2D}), because the
ultra-violet singularity at the distance of the Larmour radius
from the light cone dominates. To see this note that the
right-hand side of the analog of  (\ref{closed}) should get the
 contribution  from $D_2$:\begin{eqnarray}\label{contr2}
  \hspace{-1.5cm}\frac 1{(2\pi)^4}
\int \frac{[k_3\gamma_0-k_0\gamma_3]_{\lambda\lambda^{\rm
e}}~[k_3\gamma_0-k_0\gamma_3]_{\mu\lambda^{\rm
p}}}{(k_0^2-k_3^2)}~\frac{\exp \,[{\rm i}(kx)]}{k^2+\kappa_2}~{\rm
d}^4k\,.
\end{eqnarray} After this is contracted with the unit matrix we
get for the corresponding contribution into the right-hand side of
the equation to be written in place of Eq. (\ref{2D}) the
expression\begin{eqnarray}\label{contrunit}
  \hspace{-1.5cm}-\frac 1{(2\pi)^4}
\int ~\frac{\exp \,[{\rm i}(kx)]}{k^2+\kappa_2}~{\rm d}^4k.
\end{eqnarray}
 Once the
Hermite functions (\ref{parab1}) restrict $x_\perp$ in integrals
like (\ref{rhs3}) and (\ref{rhs4}) to the region inside the Larmour
radius, the region $k^2_\perp\gg L_{\rm B}^{-2}=m^2B/B_0$ in the
integral (\ref{contrunit}) is
%most
important. There, however, $k_2$ disappears due to the exponential
factor in (\ref{2}) and we are left with the contribution, the
same as the one coming from the free photon propagator. Moreover,
as the light-cone singularity is formed exclusively due to
integration over near-infinite values of all the four  photon
momentum components, Eq. (\ref{contrunit})  behaves like $1/x^2$,
%where $\beta$ is a constant less, $\beta < 1$,
the same as
(\ref{photon1}), near the light cone $x^2=0$.
%This remark is sufficient for
%claiming that the main contribution of $D_2$ is not suppressed for
%high magnetic field. Here we contradict the authors of Ref.
%\cite{miransky} who state that the effect of the vacuum
%polarization on the Bethe-Salpeter equation (for the Goldstonion
%in their context) reduces to the formal replacement
%$\alpha\rightarrow\alpha/2$. The point is that in Ref.
%\cite{miransky} the contribution of $D_2$ is completely
%disregarded for the reason that the term (\ref{kappa2}), linearly
%growing with the magnetic field, is in the denominator of $D_2$.
%We have seen that that this cannot be done.

We need, however, to also estimate the contribution immediately
close to this singularity. To this end, let us disregard the
spatial dispersion of the dielectric constant in the transverse
plane, i.e. take $\kappa_2$ at the value $k_\perp=0$ in
(\ref{contrunit}). By doing so we essentially underestimate the
contribution of the mode-2 photon as a carrier of the
electromagnetic interaction into the attraction force between the
electron and positron near the light cone, because we keep the
term linearly growing with the field in the denominator for large
$k_\perp$, where it in fact disappears. This approximation does
not affect the light-cone singularity, which remains $1/x^2$, but
makes the screening correction to the singular part larger than it
is. We shall see, nevertheless, that even within this
approximation, with the screening overestimated, the effect of the
latter is small. Once our working domain is restricted to the
intervals $z^2-t^2$ much closer to the light cone than the Compton
length, we may keep to the condition $|k_0^2-k_3^2|\gg m^2$ in the
integral (\ref{contrunit}). Then $\kappa_2$ (\ref{2}) should be
taken in (\ref{contrunit}) as
\begin{eqnarray}\label{kappa2}\kappa_2
=-\frac{2\alpha Bm^2}{\pi B_0}%\exp \left(-
%\frac{k_\bot^2}{2m^2}\frac{B_0}{B} \right)
.\end{eqnarray} Then (\ref{contrunit}) becomes the well-known
expression for the free propagator of a massive particle with the
mass squared $M^2=-\kappa_2$. (To avoid a possible
misunderstanding, stress that this mass should not be referred to
as an effective photon mass. The  mass of the photon defined as
its rest energy is always equal to zero, corresponding to the fact
that the point $k_0=k_3=k_1=k_2=0$ is solution to the dispersion
equations $k^2+\kappa_a(k)=0$. The "mass" $M$ appears in the
ultraviolet, and not infrared regime.) In the adiabatic
approximation of Subsection C of Section II the dimensional
reduction yields again the prescription to disregard the
dependence on  $x_\perp$ in it by setting $x_\perp=0$, $x^2=-s^2$.
Then  the contribution from (\ref{contrunit}) is
\begin{eqnarray}\label{massive}\frac{{\rm
i}M}{4\pi^2s}\;\mathcal{K}_1(Ms)\simeq\frac{\rm
i}{4\pi^2s^2}\left(1-\frac{s^2M^2}{2}\left|\ln\frac{Ms}{2}\right|\right).
\end{eqnarray} Here $\mathcal{K}_1$ is the McDonald function of order one,
and we have pointed its asymptotic behavior near the point
$s^2=z^2-t^2 = + 0~.~$ According to (\ref{spectre4}) near the
lower edge of the normalization volume and with the magnetic field
(\ref{final}) the quantity $sM=(2\alpha/\pi)^{1/2}$ makes 0.068,
and hence the second term inside the brackets in (\ref{massive})
is only $-7.8 \times 10^{-3}$. Therefore the screening effect,
although overestimated, is still negligible, the contribution of
$D_2$ making one half of the full contribution of the free photon
propagator considered above. The one half originates from the
absence of the factor 2 that appeared above when
$[\gamma_i]_{\lambda \lambda^{\rm e}}[\gamma_i]_{\mu \lambda^{\rm
p}}$ in (\ref{closed}) was later contracted with the unity $I$ to
lead to (\ref{2D}): $\sum_{i=0,3}g_{ii}\gamma_i\gamma_i=2$. The
other half comes from the contribution of $D_1$ and $D_4$.

The quantity $D_3$ contains only $k_\perp$ components that give
rise to $[k^i\gamma_i]_{\lambda\lambda^{\rm
e}}[k^j\gamma_j]_{\mu\lambda^{\rm p}}$, $i,j=1,2,$ in an equation
to appear in place of (\ref{rhs4}), and consequently contribute
only to the nondiagonal in Landau quantum numbers part of the
Bethe-Salpeter equation like (\ref{nondiag1}), that does not
survive in the limit of high magnetic field. On the contrary, the
contributions of $D_1$ and $D_4$ go to the diagonal part. This
occurs because these contain the components $k_0$ and $k_3$
carrying the matrices $\gamma_0$ and $\gamma_3$ that may lead to
the term diagonal in Landau quantum numbers, as explained when
passing from (\ref{rhs5}) to (\ref{chain}) and (\ref{closed}). It
follows from (\ref{photon2}), (\ref{b}), (\ref{D_4}) that the
common contribution from $D_1$ and $D_4$ in the (0,3) subspace is
determined by the
expression\begin{eqnarray}\label{long}\frac{(k_\perp^2)^2
k_ik_j}{(b^{(1)})^2}+\frac{k_ik_j}{k^2}=\frac{k_ik_j}{k_0^2-k_3^2},\quad
i,j=0,3.
\end{eqnarray} Then the counterpart of (\ref{contr2}) reads
\begin{eqnarray}\label{contr14}
  \hspace{-1.5cm}-\frac {1}{(2\pi)^4}
\int \frac{[k_0\gamma_0-k_3\gamma_3]_{\lambda\lambda^{\rm
e}}~[k_0\gamma_0-k_3\gamma_3]_{\mu\lambda^{\rm
p}}}{(k_0^2-k_3^2)}~\frac{\exp \,[{\rm i}(kx)]}{k^2+\kappa_1}~{\rm
d}^4k,
\end{eqnarray}and the counterpart of (\ref{contrunit}) becomes
(again with the disregard of the spatial dispersion across the
magnetic field already done when writing Eq. (\ref{1}))
\begin{eqnarray}\label{contrunit14}
  \hspace{-1.5cm}-\frac 1{\mathcal{A}(2\pi)^4}
\int ~\frac{\exp\,[{\rm i}(kx)]}{k^2}~{\rm d}^4k=\frac
1{\mathcal{A}{\rm i}4\pi^2x^2}\,,
\end{eqnarray}
where 
\begin{equation}
\mathcal{A}=1-\frac \alpha{3\pi}\left(\ln\frac
B{B_0}-C-1.21\right)
\label{mathcalA}
\end{equation} 
in view of (\ref{1}). For the fields as
large as $B=B_{\rm ult}$ (\ref{final}) the number
$\mathcal{A}$ is very close to unity: $\mathcal{A}=1-0.04.$ (Its
difference from unity is the measure of the anti-screening effect
of the running coupling constant $\alpha /\mathcal{A}$ for large
magnetic field due to the lack of
 asymptotic freedom in pure quantum electrodynamics).

 We conclude that the vacuum polarization does not any
 essentially affect the  falling to the center  and hence
 the estimate of the ultimate magnetic field. This contradicts the
 prescription to replace $\alpha\rightarrow\alpha/2$ in the
 expression for the latter that would result if one
applied the corresponding conclusion from Ref. \cite{miransky} to
the problem under consideration. The point is that in Ref.
\cite{miransky} the contribution of $D_2$ is completely
disregarded for the reason that the term (\ref{kappa2}), linearly
growing with the magnetic field, is in the denominator of $D_2$.
We saw above that that this cannot be done: it essentially
contributes to the falling to the center asymptotic regime of
$s\gg 10^{-11}m^{-1}$, where the probability to find the system is
concentrated.

Gathering the results of the present consideration together we
 conclude that the effect of the vacuum polarization leads, in the
 approximation where the spatial dispersion in the orthogonal direction
 is neglected, to the
 replacement of Eq. (\ref{closedE}) by the following two-dimensional
  Bethe-Salpeter equation for high magnetic field limit including an
  external electric field and the effects of the vacuum polarization
  \begin{eqnarray}\label{closedEV}
 \hspace{-2.5cm} \left[{\rm i}{\hat{
\partial_\|}^{\rm e}}- e\hat{A_\|}(x_\|^{\rm
e})-m\right]_{\lambda\beta} \left[{\rm i}{\hat{
\partial_\|}^{\rm p}}+e\hat{A}_\|(x^{\rm p}_\|)-m\right]_{\mu\nu}
\begin{tabular}{c}
 {\Large$[
 \chi$}$_{0,p_1^{\rm e};0,p_1^{\rm p}}(x^{\rm e}_\|, x^{\rm p}_\|)
${\Large ]}$_{\beta\nu}$\end{tabular}\nonumber\\\hspace{-2cm}
=\frac{\alpha}{\pi} ~\left\{\frac{[{\rm i}\hat{\partial}_\parallel
]_{\lambda\beta}~[{\rm i}\hat{\partial}_\parallel ]
_{\mu\nu}}{\mathcal{A}\Box_2}\frac 1{z^2+
\frac{P_1^2}{(eB)^2}-t^2}\right.+\nonumber\\
\left.\frac{[\gamma_0{\rm i}\partial_z+{\gamma_3\rm
i}\partial_t]_{\lambda\beta}~[\gamma_0{\rm
i}\partial_z+{\gamma_3\rm i}\partial_t]
_{\mu\nu}}{\Box_2}\frac{M\mathcal{K}_1\left(M(z^2+
\frac{P_1^2}{(eB)^2}-t^2)^\frac 1{2}\right)}{(z^2+
\frac{P_1^2}{(eB)^2}-t^2)^\frac 1{2}}\right\}
\begin{tabular}{c}{\Large$[
 \chi$}$_{0,p_1^{\rm e};0,p_1^{\rm p}}(x^{\rm e}_\|, x^{\rm p}_\|)
${\Large ]}$_{\beta\nu}$\end{tabular}, \end{eqnarray}Here the
action of the derivatives over $t$ and $z$ does not extend beyond
the braces, ${\rm i}\hat{\partial}_\parallel= {\rm
i}\gamma_0\partial_t+{\rm i}\gamma_3\partial_z$,
$\Box_2=\partial^2_z-\partial^2_t$. Remind that $t=x_0^{\rm
e}-x_0^{\rm p}$, $z=x_3^{\rm e}-x_3^{\rm p}.$ For no electric
field case the equation that follows from (\ref{closedEV}) for the
singlet component to substitute for (\ref{2D}) is
\begin{eqnarray}\label{2DV}\left( -\Box_2 + m^2\right) \Phi(t,z)=
\frac{2\alpha}{\pi}\left\{\frac
1{\mathcal{A}(z^2+\frac{P_1^2}{(eB)^2}-t^2)}+\frac{M\mathcal{K}_1\left(M(z^2+
\frac{P_1^2}{(eB)^2}-t^2)^\frac 1{2}\right)}{(z^2+
\frac{P_1^2}{(eB)^2}-t^2)^\frac 1{2}}\quad\right\}\Phi(t,z)
 \quad \end{eqnarray}

Finally, the Bessel equation (\ref{last2}) for the (1,1)
rotationally invariant solution now becomes
\begin{eqnarray}\label{nonbes} -\frac{{\rm d}^2\Phi}{{\rm d}
s^2}-\frac 1{s}\frac{{\rm d}\Phi}{{\rm d}
s}+m^2\Phi=\frac{2\alpha}{\pi s}\left(\frac
1{s}+M\mathcal{K}_1(Ms)\right) \Phi.\quad
\end{eqnarray} We neglected the difference of $\mathcal{A}$ from
unity.
 
\subsubsection{Mass corrections}
Mass radiative corrections should be taken into account by
inserting the mass operator into the Dirac differential operators
in the l.-h. sides of the Bethe-Salpeter equation
 ~(\ref{equation}) or (\ref{closed}). We
shall estimate now, whether this may affect the above conclusions
concerning the positronium mass compensation by the mass defect.

In strong magnetic field the one-loop calculation of the electron
mass operator leads to the so-called double-logarithm mass
correction growing with the field $B$ as \cite{jancovici}
\begin{eqnarray}
\label{mass}\widetilde{m}=m\left(1+\frac\alpha{4\pi}\ln^2\frac{B}{B_0}\right).
\end{eqnarray}
 For $B\simeq B_{\rm ult}$ the corrected mass  makes $\widetilde{m}=
3.45 m$. This
 implies that the mass annihilation
 due to the falling to the center is opposed by the radiative corrections and requires
  a field somewhat larger than (\ref{final}). To determine its value, substitute $\widetilde{m}$
(\ref{mass}) for $m$ and $L_B=(eB)^{-1/2}$ for $s_0$ into equation
(\ref{spectre3}) with $n=1$. The resulting equation for the
 ultimate magnetic field, modified by the mass radiative
 corrections, $B_{\rm corr}$,\begin{eqnarray}\label{corr}\left(1+\frac\alpha{4\pi}
 \ln^2\frac{B_{\rm corr}}{B_0}\right)^2=4\frac{B_{\rm
corr}}{B_0}\exp\left(-{\sqrt{\frac{\pi^3}{\alpha}}}+C_{\rm
E}\right),\end{eqnarray} has the numerical solution: $B_{\rm
corr}\simeq 13~ B_{\rm ult}$.

 When going beyond the one-loop
approximation by summing the rainbow diagrams two different
expressions for $\widetilde{m}$ were obtained by different
authors. Ref. \cite{loskutov2} reports\begin{eqnarray}
\label{massL}\widetilde{m}=m\exp\left(\frac\alpha{4\pi}\ln^2\frac{B}{B_0}\right).
\end{eqnarray} The use of this formula analogous to the above
 gives rise to an increase of the ultimate value by two
orders of magnitude: $B_{\rm corr}=3.5\times 10^2 B_{\rm ult}$,
whereas the use of the result of Ref.
\cite{smilga}\begin{eqnarray} \label{massS}\widetilde{m}=\frac
m{\cos\left(\sqrt{\frac\alpha{2\pi}}\ln\frac{B}{B_0}\right)}
\end{eqnarray} would leave the ultimate value practically
unchanged: $B_{\rm corr}=1.5 B_{\rm ult}$. Finally, if the vacuum
polarization is taken into account while summing the leading
contributions to the large field asymptotic behavior of the mass
operator, the following result \cite{osipov}\begin{eqnarray}
\label{massK}\widetilde{m}=\frac
m{1-\frac\alpha{2\pi}\left(\ln\frac \pi{\alpha}-C_{\rm E}\right)
\ln\frac{B}{B_0}}
\end{eqnarray} was obtained, from where the double logarithm is
absent due to the effect of the term (\ref{kappa2}) in the photon
propagator when substituted into electron-photon loops.
%$B_{\rm corr}=4.7\times 10^{28} B_0$.%
%Then, the $B$-containing contribution may become essential no
%sooner than the field is of the order of $10^{69}B_0$, which is
%much larger than (\ref{final}).
 The use of (\ref{massK}) would
 result in: $B_{\rm corr}=3B_{\rm ult}$.

Anyway, we see that the mass correction,  increasing the
ultimate value $B_{\rm ult}$ by at the most two orders of
magnitude, is not essential bearing in mind the huge values
(\ref{final}) of the latter. Moreover, basing on the most recent
results concerning the mass correction \cite{osipov} we conclude
that the latter do not affect the value of the hypercritival field
obtained above (\ref{final}) practically at all.

\section{Summary and discussion}
In the paper we have considered the system of two charged
relativistic particles - especially the electron and positron - in
interaction with each other, when placed in a strong constant and
homogeneous magnetic field ${\bf B}$. The Bethe-Salpeter equation
in the ladder approximation in the Feynman gauge is used without
exploiting any non-relativistic assumption . We have derived the
ultimate two-dimensional form of the Bethe-Salpeter equation, when
the magnetic field tends to infinity, with the help of expansion
over the complete set of Ritus matrix eigenfunctions \cite{ritus}.
The latter accumulate the spacial and spinor dependence on the
transversal-to-the-field degree of freedom. The Fourier-Ritus
transform of the Bethe-Salpeter amplitude obeys an infinite chain
of coupled differential equations that decouple in the limit of
large $B$, so that we are left with one closed equation for the
amplitude component with the Landau quantum numbers of the
electron and positron both equal to zero, while the components
with other values of Landau quantum numbers vanish in this limit.
The resulting equation is a differential equation with respect to
two variables that are the differences of the particle
coordinates: along the time $t=x_0^{\rm e}-x_0^{\rm p}$ and along
the magnetic field $z=x_3^{\rm e}-x_3^{\rm p}$. It contains only
two Dirac matrices $\gamma_0$ and $\gamma_3$ and can be
alternatively written using $2\times 2$ Pauli matrices. The term
responsible for interaction with a moderate electric field $\bf E$
directed along ${\bf B}$, $E\ll B$, is also included and does not
lay obstacles to the dimensional reduction. By introducing
different masses the resulting two-dimensional equation may be
easily modified to cover also the case of an one-electron atom in
strong magnetic field and/or other pairs of charged particles.

It is worth noting that the two-dimensionality holds only with
respect to the degrees of freedom of charged particles, while the
photons remain 4-dimensional in the sense that  the singularity of
the photon propagator is determined by the inverse d'Alamber
operator in the 4-dimensional, and not two-dimensional Minkowsky
space. (Otherwise it would be weaker).

We have made sure that in the case under consideration the
critical value of the coupling constant is zero, $\alpha_{\rm
cr}=0$, i.e., the falling to the center caused by the ultraviolet
singularity of the photon propagator as a carrier of the
interaction is present already for its genuine value
$\alpha=1/137$, in contrast to the no-magnetic-field case, where
$\alpha_{\rm cr}>1/137$. If the magnetic field is large, but
finite, the dimensional reduction holds everywhere except a small
neighborhood of the singular point $s=0$, wherein the mutual
interaction between the particles dominates over their interaction
with the magnetic field. The dimensionality of the space-time in
this neighborhood remains to be 4, and its  size is determined by
the Larmour radius $L_B=(eB)^{-1/2}$ that is zero in the limit
$B=\infty$. The latter supplies the singular problem with a
regularizing length. The larger the magnetic field, the smaller
the regularizing length, and the deeper the level.

We have found the ultimate magnetic field that provides the
full compensation of the positronium rest mass by the binding
energy, and the wave function of the corresponding state as a
solution to the Bethe-Salpeter equation. This state is described
in terms of the theory of the falling to the center, developed in
\cite{shabad,shabad2}, as a "confined" state, different from the
usual bound state. The appeal to this theory is necessitated by
the fact that the falling to the center draws the electron and
positron so close together that the mutual field is so large that
the standard treatment may become inadequate. The ultimate
value is estimated to be unaffected by the radiative corrections
modifying the mass and polarization operators.

In spite of the huge value, expected to be present, perhaps, only
in superconducting cosmic strings \cite{W85}, the magnetic
field magnitude obtained may be important as setting the limits of
applicability of QED or presenting the ultimate value of the
magnetic field admissible within pure QED. The point is that at
this field the energy gap separating the electron-positron system
from the vacuum disappears. An exceeding of the ultimate
magnetic field would cause restructuring of the vacuum. The
question about  the vacuum restructuring typical of other problems
- with or without the magnetic field, where the
falling-to-the-center takes place: the supercharged nucleus
\cite{popov,greiner} and a moderately charged nucleus with strong
magnetic field \cite{semikoz}, is discussed in the two adjacent
papers \cite{SU05,SU06}. The formal mechanisms that realize the
magnetic field instability and may lead to prevention of its
further growth via the decay of the "confined" state found here
require a further study and will be considered elsewhere.

\begin{acknowledgments}The authors are grateful to V.~Gusynin,
V.~Miransky and N.~Mikheev for valuable discussions. This work was
supported by the Russian Foundation for Basic Research (project no
05-02-17217) and the President of Russia Programme
(LSS-1578.2003.2), as well as by the Israel Science Foundation of
the Israel Academy of Sciences and Humanities.
\end{acknowledgments}

{}

\begin{thebibliography}{}
\bibitem{footnote2} {Henceforth, if superscripts ~e~ or ~p~ are
omitted, the corresponding equations relate both to electrons and
positrons in a form-invariant way.}
\end{thebibliography}
\end{document}